\documentclass[fleqn,usenatbib]{mnras}

\usepackage{newtxtext,newtxmath}

\usepackage[T1]{fontenc}

\DeclareRobustCommand{\VAN}[3]{#2}
\let\VANthebibliography\thebibliography
\def\thebibliography{\DeclareRobustCommand{\VAN}[3]{##3}\VANthebibliography}

\usepackage{graphicx}	
\usepackage{amsmath}	
\usepackage{ragged2e}
\usepackage{longtable}
\usepackage{rotating} 

\providecommand{\teff}{\ensuremath{T_{\rm eff}}}
\providecommand{\teffsun}{\ensuremath{T_{{\rm eff},\odot}}}

\providecommand{\msun}{\ensuremath{\,M_{\odot}}}
\providecommand{\rsun}{\ensuremath{\,R_{\odot}}}

\providecommand{\mj}{\ensuremath{\,M_{\rm J}}}

\newcommand{\tess}{\emph{TESS}}
\newcommand{\ms}{\mbox{m\,s$^{-1}$}}
\newcommand{\Lsun}{\mbox{L$_{\odot}$}}

\newcommand{\kms}{\,km\,s$^{-1}$}

\newcommand\msyr{\ensuremath{\text{m}\,\text{s}^{-1}\,\text{yr}^{-1}}}
\newcommand\masyr{\ensuremath{\text{mas}\,\text{yr}^{-1}}}

\usepackage{CJKutf8}
\newcommand{\CNnames}[1]{{\begin{CJK}{UTF8}{gbsn}~(#1)~\end{CJK}}}

\newcommand{\numax}{\ensuremath{\nu_{\textrm{max}}}}
\newcommand{\numaxsun}{\ensuremath{\nu_{\textrm{max},\odot}}}
\newcommand{\Dnu}{\ensuremath{\Delta\nu}}
\newcommand{\Dnusun}{\ensuremath{\Delta\nu_{\odot}}}
\newcommand{\echelle}{\'{e}chelle}
\newcommand{\muHz}{\mbox{$\mu$Hz}}

\title[PPPS IX.]{The Pan-Pacific Planet Search -- IX. A menagerie of companions orbiting evolved stars}

\author[R.A. Wittenmyer et al.]{
Robert A. Wittenmyer,$^{1}$\thanks{E-mail: rob.w@usq.edu.au}
Alexander Venner,$^{2,1}$
Tyler Fairnington,$^{1,3}$
George Zhou,$^{1}$
\newauthor Duncan J. Wright,$^{1}$
Evan Curtin,$^{4}$
Timothy R. Bedding,$^{5}$
Courtney L. Crawford,$^{5}$
Yaguang Li\CNnames{李亚光},$^{6}$
\newauthor Dennis Stello,$^{7}$
Marc Hon,$^{8}$
Daniel Huber,$^{6}$
Frank Grundahl,$^{9}$
M. Skakke Fredslund,$^{9}$
Pere L. Palle,$^{16,17}$
\newauthor Tianjun~Gan,$^{10}$
Jonathan Horner,$^{1}$
John Kielkopf,$^{11}$
Stephen R. Kane,$^{12}$
Peter Plavchan,$^{13}$
\newauthor Avi Shporer,$^{14}$
C.G. Tinney,$^{7}$
Hui Zhang,$^{15}$
Matthew W. Mengel,$^{1}$
Jack Okumura$^{1}$
\\
$^{1}$University of Southern Queensland, Centre for Astrophysics, USQ Toowoomba, QLD 4350 Australia\\
$^{2}$Max Planck Institute for Astronomy, 69117 Heidelberg, Germany\\
$^{3}$Department of Astronomy \& Astrophysics, University of Chicago, Chicago USA \\
$^{4}$Department of Physical Sciences, Kutztown University, Kutztown, PA 19530, USA \\
$^{5}$Sydney Institute for Astronomy (SIfA), School of Physics, University of Sydney, NSW 2006, Australia\\
$^{6}$Institute for Astronomy, University of Hawai`i, 2680 Woodlawn Drive, Honolulu, HI 96822, USA\\
$^{7}$School of Physics, UNSW Sydney, NSW 2052, Australia\\
$^{8}$Department of Physics, National University of Singapore, 21 Lower Kent Ridge Road, Singapore, 119077\\
$^{9}$Stellar Astrophysics Centre (SAC), Department of Physics and Astronomy, Aarhus University, Ny Munkegade 120, DK-8000 Aarhus, Denmark\\
$^{10}$Department of Astronomy, Westlake University, Hangzhou 310030, Zhejiang Province, China\\
$^{11}$Department of Physics and Astronomy, University of Louisville, Louisville, KY 40292, USA\\
$^{12}$Department of Earth and Planetary Sciences, University of California, Riverside, CA 92521, USA\\
$^{13}$Department of Physics \& Astronomy, George Mason University, 4400 University Drive MS 3F3, Fairfax, VA 22030, USA\\
$^{14}$Department of Physics and Kavli Institute for Astrophysics and Space Research, Massachusetts Institute of Technology, Cambridge, MA 02139, USA\\
$^{15}$Shanghai Astronomical Observatory, Chinese Academy of Sciences, Shanghai 200030, People’s Republic of China\\
$^{16}$Instituto de Astrofísica de Canarias, 38200 La Laguna, Tenerife, Spain\\
$^{17}$Departamento de Astrofísica, Universidad de La Laguna (ULL), 38206 La Laguna, Tenerife, Spain
}

\date{Accepted XXX. Received YYY; in original form ZZZ}

\pubyear{\the\year{}}

\begin{document}
\label{firstpage}
\pagerange{\pageref{firstpage}--\pageref{lastpage}}
\maketitle

\begin{abstract}
We present resolutions as to the nature of six speculative candidate companions proposed in the final data release of the Pan-Pacific Planet Search, a 6-year radial-velocity survey of 164 southern evolved stars using the now-decommissioned UCLES spectrograph on the 3.9m Anglo-Australian Telescope.  New radial-velocity observations, \textit{TESS} asteroseismology, and \textit{Hipparcos-Gaia} astrometry are incorporated to refine the companion and host-star parameters.  We confirm that HD\,126105b is a giant planet ($P=524.0\pm$2.9\,d, $m$ sin $i=1.67^{+0.19}_{-0.17}$\mj), and that HD\,205577B is a massive, eccentric brown dwarf ($P\sim$11.2\,yr, $m=77^{+11}_{-9}$\mj, $e=0.68$).  HD\,115066B and HD\,121156B are low-mass stellar companions, while HD\,114899 and HD\,159743 are shown to be unadorned by any detectable companions whatsoever.  This demonstrates the utility of astrometric information to help overcome the temporal limitations of incomplete radial-velocity data sets and elucidate the true nature of suspected companion bodies. 

\end{abstract}




\section{Introduction}

While many thousands of planets have been confirmed orbiting other stars, the vast majority of them orbit main-sequence stars.  Only 7\% of confirmed exoplanets orbit evolved stars, defined as $\log g<4.0$\footnote{Based on planet data from the NASA Exoplanet Archive, accessed on 2025 March 6.}.  The detailed demographic properties of planets orbiting evolved stars are thus far less constrained than those orbiting main-sequence stars, particularly for longer-period orbits.  A small but determined cohort of astronomers has made continuing efforts to push back the veil of time to better understand the populations and architectures of planetary systems beyond the main sequence.  

The search for planets orbiting evolved stars was borne out of the desire to understand the properties and occurrence of planets orbiting stars more massive than our Sun.  The early radial-velocity planet search programs heavily biased their target lists in favour of Sun-like main-sequence stars, resulting in a strong over-representation of host stars with masses near 1\msun.  This is a consequence of the radial-velocity technique, the precision of which requires an abundance of narrow spectral lines: the types of spectra characteristic of main-sequence FGK type stars.  As a result, more massive stars (A and early F) were cast aside as lacking sufficient spectral information to measure radial velocities useful for exoplanet detection.  Their spectral lines are simply too sparse and the stars rotate too quickly to achieve the $\sim$3-5 \ms\ precision necessary.  

Stellar astrophysics, though sometimes cruel and uncaring, can on occasion proffer a ray of hope to the dedicated astronomer.  The solution all along is to wait.  When earlier-type stars, of masses 1.5-3\msun, evolve off the main sequence, their spectroscopic intransigence gives way to an abundance of narrow lines more suited for precise radial-velocity measurements.  This is a result of their atmospheres expanding and cooling as they evolve: both effects are favourable as they allow neutral metal lines to form without the $\sim$50-100\kms\ rotational broadening exhibited on the main sequence.  So it goes that in the mid-2000s, several pioneering campaigns\footnote{Precise RV facilities used in these surveys include: Keck/HIRES, Okayama, CHIRON, FEROS, Lick, and CORALIE.} began to take advantage of this rare gift from physics to explore the population of planets around the ``retired A stars'' \citep[e.g.][]{giant1, giant2, ppps1, giant3, giant4, cascades}.  

The legacy radial-velocity sample of evolved stars, combined with a growing sample from the \textit{Transiting Exoplanet Survey Satellite (TESS)} including main-sequence intermediate-mass stars \citep[e.g.][]{zhou19, grunblatt19, saunders22}, has permitted some preliminary demographic analysis of the population of planets as a function of host-star mass \citep[e.g.][]{jones16, ppps8, wolthoff22, teng23}. In brief, the occurrence, masses, and multiplicity of planetary systems hosted by these stars is positively correlated with stellar metallicity.  The planet occurrence rate appears to peak for $\sim$1.7\msun hosts, falling off precipitously at higher stellar masses.  

Among the $\sim$400 planets confirmed to orbit these evolved stars, there are of course some intriguing individual systems.  HD\,76920b was found to be the most eccentric planet to orbit an evolved star \citep{76920, bergmann21}, with $e=0.8782\pm0.0025$.  \citet{hon23} revealed an ``impossible'' close-in planet orbiting the core helium-burning giant 8 UMi that appears to have escaped engulfment during the star's first ascent onto the red giant branch.  Longer-baseline radial-velocity observations have revealed planet pairs in low-order mean-motion resonances, such as HD\,200964 and 24 Sextantis \citep{johnson11, witt12}, HD\,33844 \citep{33844}, and 7 CMa \citep{luque19}.  Recently, \citet{teng22} reported a unique compact triple-giant-planet system orbiting HD\,184010, where the period ratios of neighbouring planet pairs are less than 2:1.  While it may appear that some planets are more equal than others, for the purposes of understanding the population, we must consider all planets of equal value, even the ``boring ones.''  Getting it right is key.  For example, \citet{ppps6} considered and rejected a planet candidate orbiting HD\,29399; it was later confirmed as a genuine planet with further data and a more rigorous stellar activity analysis by \citet{pezzotti22}.  The 1.57\mj\ planet moving on an 893-day, nearly-circular orbit is fairly typical of the planets known to orbit these evolved stars, but that makes it no less important for adding to our understanding of the overall population.  


\citet{ppps8} presented the final data release for the Pan-Pacific Planet Search \citep{ppps1}, a collaboration between Australia, China, and the US, which obtained precise radial velocities for 129 bright Southern hemisphere, evolved intermediate-mass stars from 2009 to 2015.  That work noted 12 speculative candidate companions with incomplete or uncertain orbits.  Here we consider the six of those stars for which we have obtained sufficient follow-up observations to revisit the analysis of their true natures.  This is their story.  In 2019-2020, we started follow-up observations of HD\,126105 with the Hertzsprung SONG (Stellar Observations Network Group) telescope at the Teide Observatory in Tenerife. We also began to monitor the remaining five more southerly targets with the \textsc{MINERVA}-Australis telescope array in southern Queensland, Australia.  The other six targets noted in \citet{ppps8} have insufficient data to warrant further investigation at this time.  


In this paper, we report the discovery and confirmation of four multifarious companions orbiting evolved stars: the giant planet HD\,126105b, the high-mass brown dwarf HD\,205577B, and the stellar companions HD\,115066B and HD\,121156B.  The latter two objects were first identified as candidate stellar-mass companions with uncertain, incomplete orbits in \citet{bluhm16}.  HD\,205577B is a rare highly eccentric brown dwarf in a $\sim$11 yr orbit, with a derived true mass of 77$^{+11}_{-9}$ \mj, on the borderline between a brown dwarf and extremely low-mass star.  Section 2 details the radial-velocity observations, and in Section 3, we describe the properties of the host stars.  Section 4 describes the radial-velocity and astrometric fits for these objects, and we give our results and conclusions in Section 5.


\section{Observations and Data Reduction}\label{sec:observations}

All six stars considered here were observed with AAT/UCLES as part of the Pan-Pacific Planet Search.  The radial velocity data for these and all survey stars are given in \citet{ppps8} and reprised here in the Appendix for convenience (Tables~\ref{tab:114899rv}-\ref{tab:205577rv_2}).  \citet{bluhm16} also presented FEROS and CHIRON data from the EXPRESS survey \citep{jones11} for HD\,115066 and HD\,121156, which have been included in the fitting process as described in Section \ref{sec:Fitting}.  

\subsection{SONG}

We obtained 13 radial velocity measurements of HD\,126105 with the robotic 1m Hertzsprung SONG telescope located at the Teide Observatory in Tenerife \citep{2007CoAst.150..300G, frank2017, Andersen_2019}. Light from the telescope is guided to the slit-based high-resolution echelle spectrograph using a coudé path. Slit number 6 was used which provides a resolution of $R=90,000$. HD\,126105 was observed between 2019 Dec 13 and 2022 July 14 with an exposure time of 40 minutes. Each spectrum covers 440 – 690\,nm and the stellar light passes through a heated iodine cell for precise wavelength calibration. The radial velocities were extracted from the spectra using the \texttt{pyodine} software described by \citet{heeren23}.  

\subsection{MINERVA-Australis}

We carried out supplemental spectroscopic observations using the {\textsc{Minerva}}-Australis facility \citep{addison2019, toi1842, clark23}.  {\sc {\textsc{Minerva}}}-Australis consists of an array of four independently operated 0.7\,m CDK700 telescopes situated at the Mount Kent Observatory in Queensland, Australia \citep{addison2019}.  Each telescope simultaneously feeds stellar light via fibre optic cables to a single KiwiSpec R4-100 high-resolution ($R=80,000$) spectrograph \citep{2012SPIE.8446E..88B} with wavelength coverage from 480 to 620\,nm.  Wavelength calibration is achieved via archival Th-Ar wavelength solution. The wavelength solution is corrected for each exposure via a simultaneous iodine gas cell calibration fibre. The calibration fibre is illuminated by a quartz flat lamp, with the light passing through an iodine gas cell to provide the absorption lines for wavelength corrections. 

For HD\,126105, we obtained 17 epochs from 2024 March 10 to 2024 July 27.  For HD\,126105 we obtained 12 new RVs from 2023 April 13 to 2023 August 19.  For HD\,121156 we acquired 29 RVs between 2023 February 27 to 2024 September 4.  For HD\,205577, we obtained 13 epochs from 2021 April 9 to 2022 August 24, and a further 47 epochs from 2023 May 3 to 2024 July 14 after a fibre switch that took place in 2022 December.  These are treated as two separate instruments in the fitting process, labelled ``pre'' and ``post'' hereafter.  Exposure times were 30 minutes, distributed amongst up to four {\textsc{Minerva}}-Australis telescopes.  Radial velocities were derived for each telescope by cross-correlation, where the template being matched is the mean spectrum of each telescope.  Radial velocities from multiple telescopes at a given epoch were binned into a single point.

\section{Stellar Properties}\label{sec:thestar}

Complete stellar parameters for the six stars considered in this work are given in Table~\ref{tab:allstars}.  We present updated stellar parameters for HD\,126105 incorporating new \textsc{MINERVA}-Australis spectra and \textit{Gaia} DR3 information.  For HD\,115066 and HD\,205577, we have derived stellar parameters from a new asteroseismic analysis of high-precision photometry from NASA's \textit{Transiting Exoplanet Survey Satellite} (TESS) \citep{tess}. 


\begin{table*} 

\label{stellarprops}

\caption{Stellar properties for host stars of the systems examined in this study.  Notes: Adopted values are shown in bold.} 

\label{tab:allstars}      
\centering                                
\rotatebox{90}{%
\begin{tabular}{l cl cl cl cl cl cl }        

\hline\hline                 

Property & HD\,114899 & Ref. & HD\,115066 & Ref. & HD\,121156  & Ref. & HD\,126105 & Ref. & HD\,159743 & Ref. & HD\,205577 & Ref. \\  
\hline 

Right Ascension (h:m:s)   & 13:14:26.24 & 1       & 13:15:04.32 & 1               &  13:54:16.64 & 1              & 14:24:00.36 & 1                      & 17:37:01.63 & 1            & 21:36:43.65 & 1  \\
Declination (d:m:s)       & -54:57:43.69 & 1      & -30:10:53.07 & 1              & -28:34:10.62 & 1              & -19:48:03.18 & 1                     & -18:59:31.48 & 1           & -21:30:10.22 & 1  \\
Distance (pc)             & 199.09$\pm$0.94  & 1  & 187.4~$\pm$~1.05 & 1          &  63.899~$\pm$~0.16  & 1       & 82.32~$\pm$~0.18 & 1                 & 132.41$\pm$0.41 & 1        & 210.8~$\pm$~1.2 & 1  \\
Spectral type             & K0 III & 2            & K0 II & 2                     &  K2 III & 2                   &  K1 III & 2                          & K0 III & 2                 & K0/1 III & 2  \\
$V$ (mag)                 & 7.98$\pm$0.01 & 3     & 7.82~$\pm$~0.01 & 3           &  6.045~$\pm$~0.010 & 3        & 7.34~$\pm$~0.01 & 3                  & 7.45$\pm$0.01 &  3         & 7.93~$\pm$~0.01 & 3  \\
$G$ (mag)                 & 7.711$\pm$0.003 & 1   & 7.534~$\pm$~0.003 & 1         & 5.737~$\pm$~0.003  & 1        & 7.066~$\pm$~0.003  & 1               & 7.149$\pm$0.003 & 1        & 7.604~$\pm$~0.003  & 1  \\
$T_{\rm eff}$ (K)         & 4858 & 4              & 4865$\pm$60 & 7               &  \textbf{4670} & 9            & 4732 & 11                            & 4746$\pm$50 &  6           & 4614~$\pm$~100 & 14  \\
                          & 4965$\pm$100 & 5      & \textbf{4632$\pm$92} & 7      &  4486$\pm$89 & 7              &  4870~$\pm$~100 & 5                  & 4730$\pm$50 & 13           & & \\
                          &  &                    & 4752$\pm$100 & 5              &   4734  & 10                  & \textbf{4875$^{+45}_{-40}$} & 12     & 4706$\pm$100 & 5           & & \\
                          &  &                    &              &                &   4804$\pm$72 & 6             &                          &           &  &                         & & \\
                          &  &                    &              &                &    4710$\pm$100 & 5           &                          &           &  &                         & & \\
$\log g$ (cm\,s$^{-2}$)   & 3.09$\pm$0.15 & 5     & 2.72$\pm$0.14  & 6            & \textbf{2.937$\pm$0.005} & 9  & 3.3  & 11                            & 2.783$\pm$0.099 & 6        & 2.63$\pm$0.15 & 14 \\
                          &  &                    & 2.85$\pm$0.15 & 5             &   2.981  & 10                 & 3.39$\pm$0.15 & 5                    & 2.73$\pm$0.1 & 13          & & \\
                          &  &                    &              &                &   2.875$\pm$0.149 & 6         & \textbf{3.18$^{+0.12}_{-0.11}$} & 12 & 2.94$\pm$0.15 & 5          & & \\
                          &  &                    &              &                &   3.12$\pm$0.15 & 5           &                          &           &  &                         & & \\
$R_{\star}$ ($R_{\odot}$) & 6.08$\pm$0.1 & 5      & 8.0$\pm$0.1 & 5               & \textbf{6.54$\pm$0.11} & 9    & 3.89$\pm$0.1  & 5                    & 8.26$\pm$0.1 & 5           & 8.88  & 14 \\
                          &  &                    & 8.5$\pm$0.4 & 7               &  6.3$\pm$0.9 & 7              & \textbf{4.21$\pm$0.06} & 12          & 7.12$\pm$0.09 & 8          & \textbf{9.93$\pm$0.42} & 14 \\
                          &  &                    & \textbf{8.36$\pm$0.14} & 14   &  6.3$\pm$0.6 & 5              &                           &          &  &                         & & \\
$L_{\star}$ ($L_{\odot}$) & 19.1 & 5              & 30.2  & 6                     &   18.2    & 6                 & 7.6  & 5                             & 22.9 & 6                   & 33.1  & 5 \\
                          & 17.0 & 4              & \textbf{29.7$\pm$1.0} & 7     &   \textbf{14.3$\pm$3.8} & 7   & \textbf{9.0$\pm$0.4} & 12            & 30.9 & 5                   & \textbf{38.41$\pm$1.25} & 7  \\
                          &  &                    & 19.5 & 5                      &  19.5 & 5                     &                        &             &  &                         & & \\
$M_{\star}$ ($M_{\odot}$) & 1.51 &  5             & 1.36$\pm$0.11  & 6            & \textbf{1.35$\pm$0.05} & 7     & 1.08$\pm$0.25  & 5                   & 1.15$^{+0.11}_{-0.10}$ & 6 & 1.10$\pm$0.25  & 5 \\
                          &  &                    & \textbf{1.26$\pm$0.06} &  14  & 1.33$\pm$0.25 & 5             & \textbf{1.20$^{+0.14}_{-0.21}$} & 12 & 1.45$\pm$0.25 & 5          & \textbf{1.14$\pm$0.13} & 14 \\
Metallicity, [Fe/H]       & -0.01$\pm$0.10 & 5    & \textbf{-0.13$\pm$0.055} & 6  & 0.13~$\pm$~0.06 & 6           &  0.01~$\pm$~0.10 & 5                 & -0.11$\pm$0.05 & 6         & -0.14$\pm$0.10 & 5 \\
                          &  &                    & -0.18$\pm$0.10 & 5            &  0.25$\pm$0.10 & 5            & \textbf{0.00~$\pm$~0.04} & 12        & -0.20$\pm$0.10 & 5         & & \\
Age (Gyr)                 & 2.91 & 5              & 3.31  & 6                     & 3.45  & 6                     & 4.9$^{+4.6}_{-1.2}$ & 12             & 6.06 & 6                   & 6.4~$\pm$~3.3 & 14 \\
                          &  &                    & 4.2$\pm$0.7 & 14              &  4.51 & 5                     &                           &          & 3.40 & 5                   & & \\
$v \sin i$ (\kms)         &  &                    & 3.19$\pm$0.60  &  8           & 3.4$\pm$0.8  &  8             & 2.1$\pm$0.8 & 5                      & 2.7$\pm$0.5 & 8            & 5.5  &  7 \\
                          &  &                    &              &                &              &                & 2.9$\pm$2.3 & 12                     &  &                         & & \\

\noalign{\vskip 6px} 
\hline
\noalign{\vskip 8px} 
\end{tabular} }
\\
\justifying
1. \cite{GaiaDR3}, 2. \cite{MSS1982}, 3. \cite{2000Hog}, 4. \cite{mcdonald2012}, 5. \cite{witt16}, 6. \cite{aguilera23}, 7. \cite{Hon++2021}, 8. \cite{soto21}, 9. \cite{pope24}, 10. \cite{perd2024}, 11. \cite{massarotti}, 12. textsc{Minerva}-Australis spectra (this work), 13. \cite{soubiran2022}, 14. Seismic analysis (this work).  \\
  
\justifying
\end{table*}

\subsection{Spectroscopic and photometric analysis of HD\,126105}

For HD\,126105, the eight spectra with the highest SNRs were convolved into a single spectrum with a resulting SNR of approximately 133. The spectrum was binned every 0.02 Angstroms using the binningx0dt function found within PyAstronomy \footnote{\url{https://github.com/sczesla/PyAstronomy}} to reduce the influence of any outliers on the data fitting \citep{pya}. Before convolving the spectrum, all spectra were first corrected for their barycentric velocity at the time of observation and then corrected for radial velocities measured from a cross correlation with the mean spectrum of the target.  Then, utilising iSpec \citep{ispec2014, ispec2019}, the convolved spectrum was fitted with a model for effective temperature ($T_{\rm eff}$), surface gravity ($\log g$), and metallicity ([M/H]). MARCS \citep{MARCS} was used as the atmospheric model when performing standard iSpec procedures.

The astroARIADNE package was used \citep{ARIADNE} to verify $T_{\rm eff}$, $\log g$, and [M/H], while also extracting the radius ($R_{\star}$), luminosity ($L_{\star}$), and mass ($M_{\star}$) of HD\,126105. AstroARIADNE is a python package that utilizes photometric data from Gaia Data Release 3 (DR3) to fit stellar atmospheric models with nested sampling algorithms \citep{Skilling2004, Skilling2006}. AstroARIADNE also performs Bayesian model averaging (BMA) to average posterior samples from multiple stellar atmosphere models while correcting for any biases found within each model. Four models: Phoenix \citep{Husser2013}, BT-Settl \citep{Allard2012}, Kurucz \citep{1993KurCD..13.....K}, and Castelli \& Kurucz \citep{atlas9} were used in the BMA processes for HD\,126105. All fitted parameters: $T_{\rm eff}$, $\log g$, [M/H], $R_{\star}$, $L_{\star}$, and $M_{\star}$, are listed in Table \ref{tab:allstars}.

Finally, from the spectroscopic $v\sin i$ and the stellar radius we obtain an upper limit on the stellar rotation period, $P_{\rm rot}/\sin i \sim 101$~d.

\subsection{Asteroseismic analysis of HD\,205577}
\label{sec:seismo-HD205577}

\begin{figure*}
    \centering
 
     \includegraphics[width=0.8\linewidth]{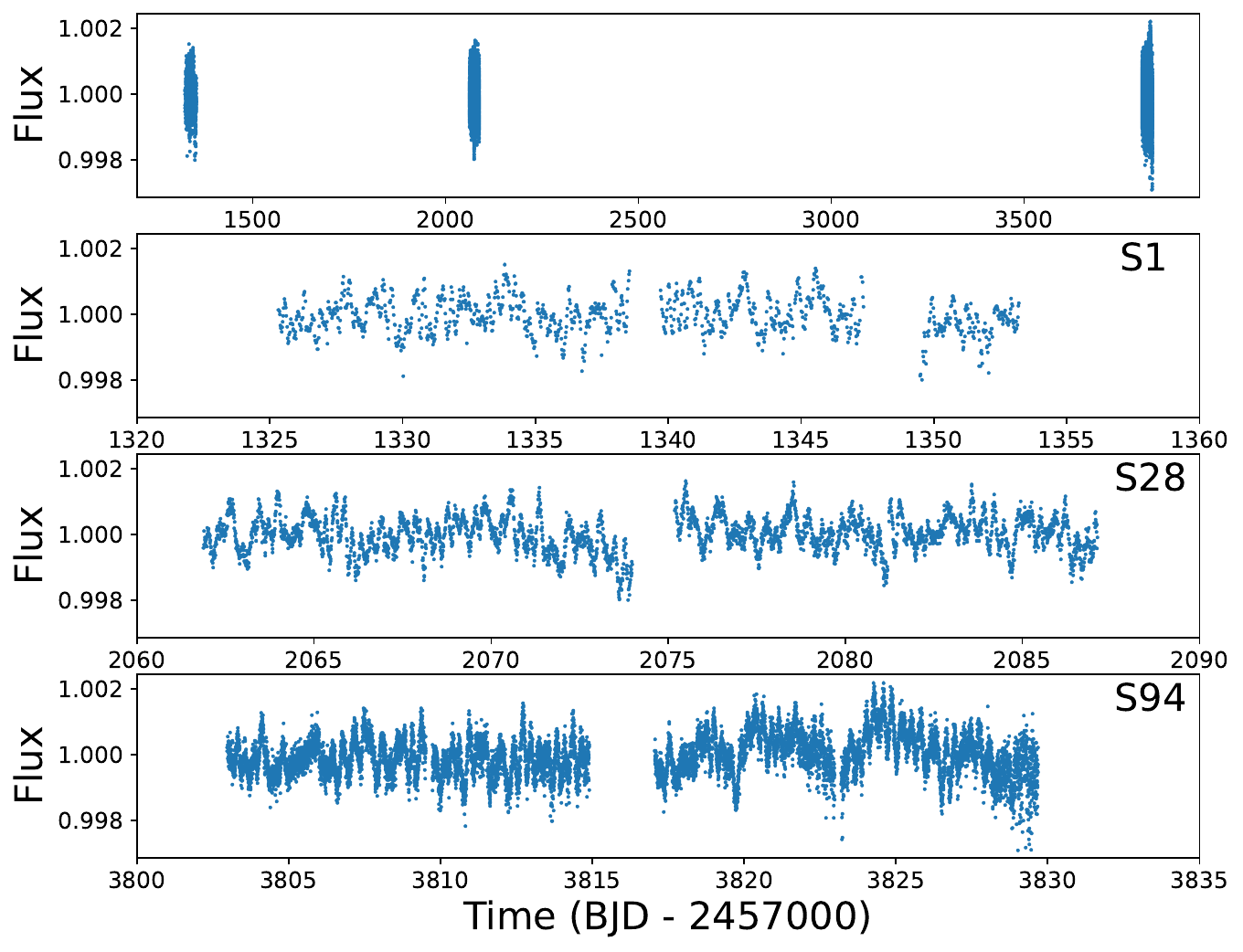}

     \caption{\textit{TESS} light curve of HD\,205577, showing the full light curve (top), Sector 1 (upper middle), Sector 28 (lower middle), and Sector 94 (bottom). 
     } 
    \label{fig:lightcurve} 
\end{figure*}

\begin{figure}
    \centering

    \includegraphics[width=\columnwidth]{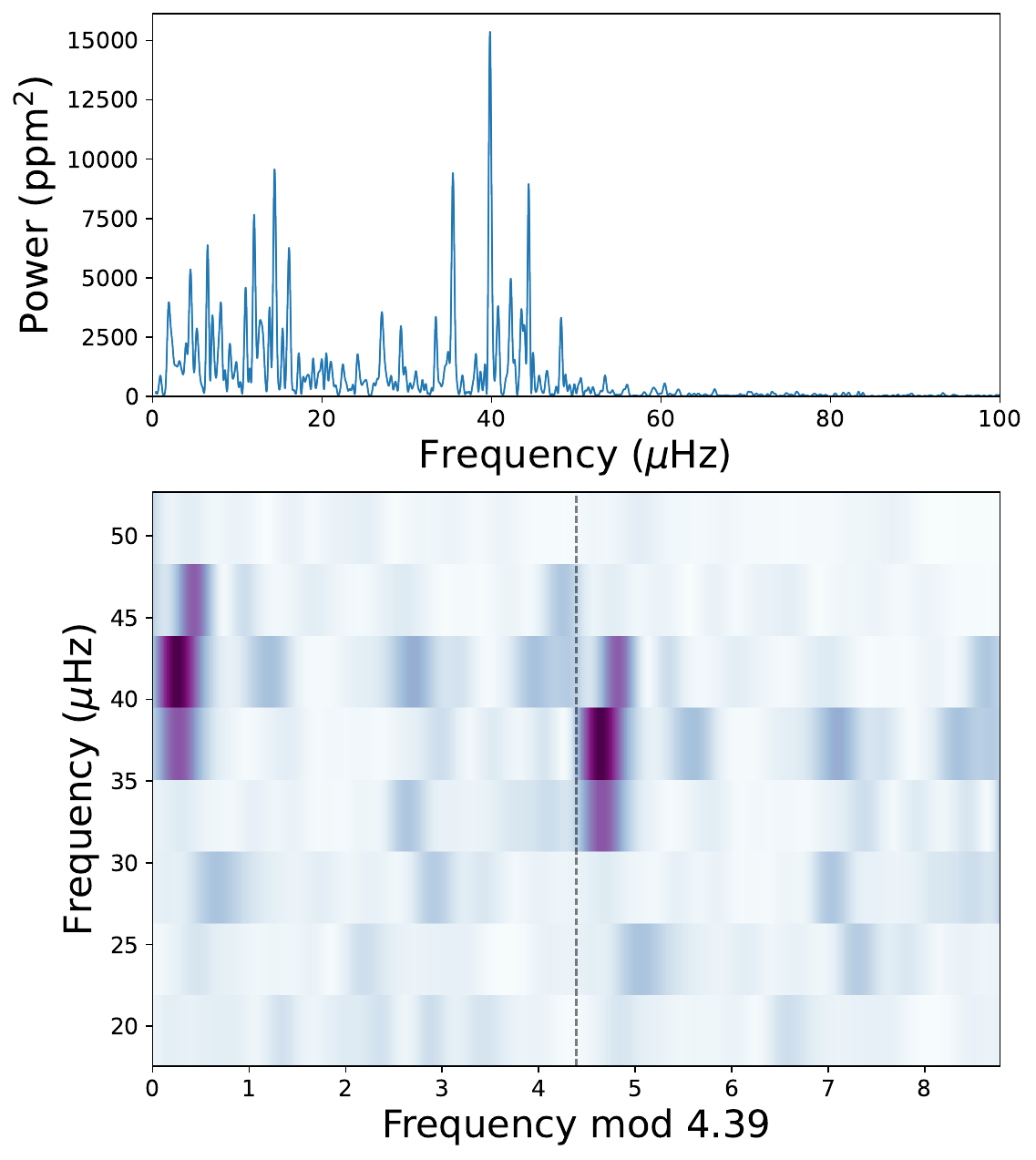}

     \caption{Upper: GLS power spectrum of the full \textit{TESS} light curve for HD\,205577, showing solar-like oscillations centred at about 40\,\muHz.  Lower: Power spectrum of the \textit{TESS} light curve of HD\,205577 in \echelle\ format (see text).
     } 
    \label{fig:power-spectrum} 
\end{figure}

Solar-like oscillations can be used to constrain the properties of red giants, such as mass, radius and age \citep[see reviews by][]{Chaplin+Miglio2013, Garcia+Ballot2019, Jackiewicz2021}.  To measure oscillations in HD\,205577, we analyzed the three available sectors of TESS data (Sectors 1, 28, and 94).  We created a light curve from the \tess\ target pixel files using the {\tt lightkurve} package \citep{lightkurve} and high-pass filtered the light curve to remove slow variations. The resulting light curve is shown in Fig.~\ref{fig:lightcurve}, and oscillations with a period of about 0.3\,d are visible.  This is confirmed by the Generalised Lomb-Scargle (GLS) power spectrum of the light curve (Fig.~\ref{fig:power-spectrum}), which shows an excess centred at $\numax \simeq 40\,\muHz$ from oscillations, as well as a rising level of power towards low frequencies that is due to granulation on the stellar surface.

The frequency \numax, which measures the centre of the oscillation envelope, scales approximately as a function of the surface gravity and effective temperature of the star \citep{Brown++1991, Kjeldsen+Bedding1995}:
\begin{equation}
    \frac{\numax}{\numaxsun} \simeq \frac{M}{\msun} \left(\frac{R}{\rsun}\right)^{-2} 
    \left(\frac{\teff}{\teffsun}\right)^{-0.5}.
\end{equation}
The value of \numax\ for HD\,205577 was measured by \citet{Hon++2021} from TESS data to be $41.4 \pm 2.1\,\muHz$, using a machine learning technique. 

Meanwhile, the so-called large separation between consecutive overtone modes scales as a function of the mean stellar density \citep{Ulrich1986}:
\begin{equation}
    \frac{\Dnu}{\Dnusun} \simeq \left(\frac{M}{\msun}\right)^{0.5} 
    \left(\frac{R}{\rsun}\right)^{-1.5}.
\end{equation}
For a red giant with the observed value of \numax, we expect \Dnu\ to be about 4--5\,\muHz\ \citep[e.g.,][]{Yu++2018}, and a regular series of peaks with this spacing is indeed apparent in the power spectrum.  To measure \Dnu\ more exactly, we constructed a so-called \echelle\ diagram, in which the power spectrum is divided into equal segments of length \Dnu\ that are stacked vertically to form an image.  When the correct value of \Dnu\ is chosen, the peaks align in vertical ridges that correspond to modes with different angular degrees \citep[for more details see, e.g.,][]{Bedding2014}.  The lower panel of Fig.~\ref{fig:power-spectrum} shows the \echelle\ diagram using a value of $\Dnu = 4.39\,\muHz$ (the \echelle\ image is plotted twice for clarity; see \citealt{Bedding2012}). We adopted an uncertainty on \Dnu\ of 0.03\,\muHz.



%
%

To infer the stellar properties from \numax\ and \Dnu, we adopted the following properties for the star:
$\teff = 4614 \pm 100$\,K \citep{witt16};
${\rm [Fe/H]} = -0.14 \pm 0.10$  \citep{witt16};
$[\alpha/{\rm Fe}] = 0.0$ (in the absence of a measurement); and
$L = 38.41 \pm 1.25 \,\Lsun$ \citep{Hon++2021}.

We fitted the observables to two grids of red giant models, namely those calculated by \citet{Stello++2009} and \citet{Li-Yaguang++2023}. The five observational constraints (\numax, \Dnu, \teff, $L$ and [Fe/H]) were used to independently construct the $\chi^2$ statistics. Stellar parameters were estimated by marginalizing the likelihood functions, which are proportional to $\exp(-\chi^2/2)$. The model fits gave the following stellar parameters for HD\,205577:
$R= 9.93 \pm 0.42\,\rsun$,
$M= 1.14 \pm 0.13\,\msun$ and 
${\rm age} = 6.4 \pm 3.3$\,Gyr.
The results given by the two model grids agreed within the reported 1-$\sigma$ uncertainties.


\subsection{Asteroseismic analysis of HD\,115066}

\begin{figure*}
    \centering
 
     \includegraphics[width=0.8\linewidth]{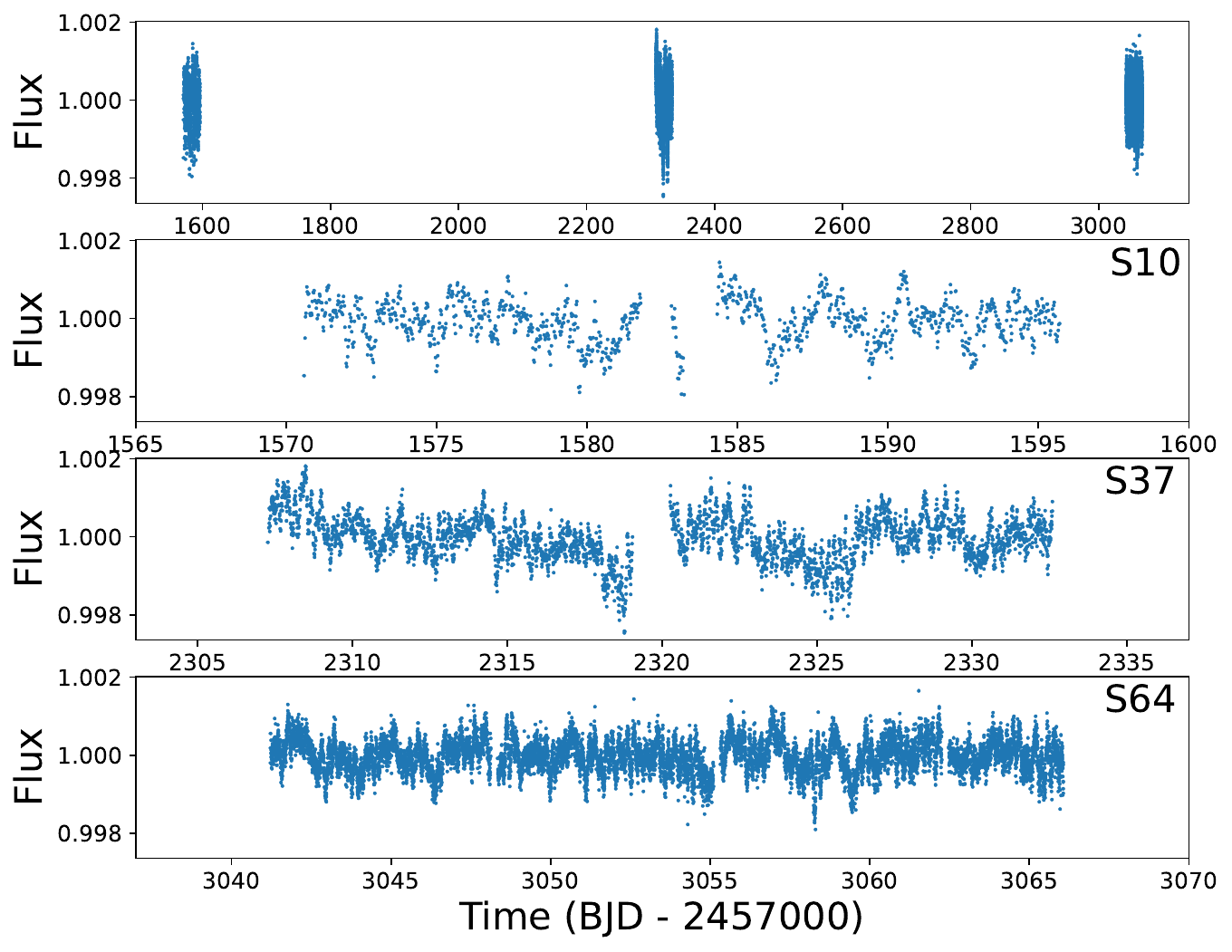}

     \caption{\textit{TESS} light curve of HD\,115066, showing the full light curve (top), Sector 10 (upper middle), Sector 37 (lower middle), and Sector 64 (bottom). 
     } 
    \label{fig:HD115066_lightcurve} 
\end{figure*}

\begin{figure}
    \centering

     \includegraphics[width=\columnwidth]{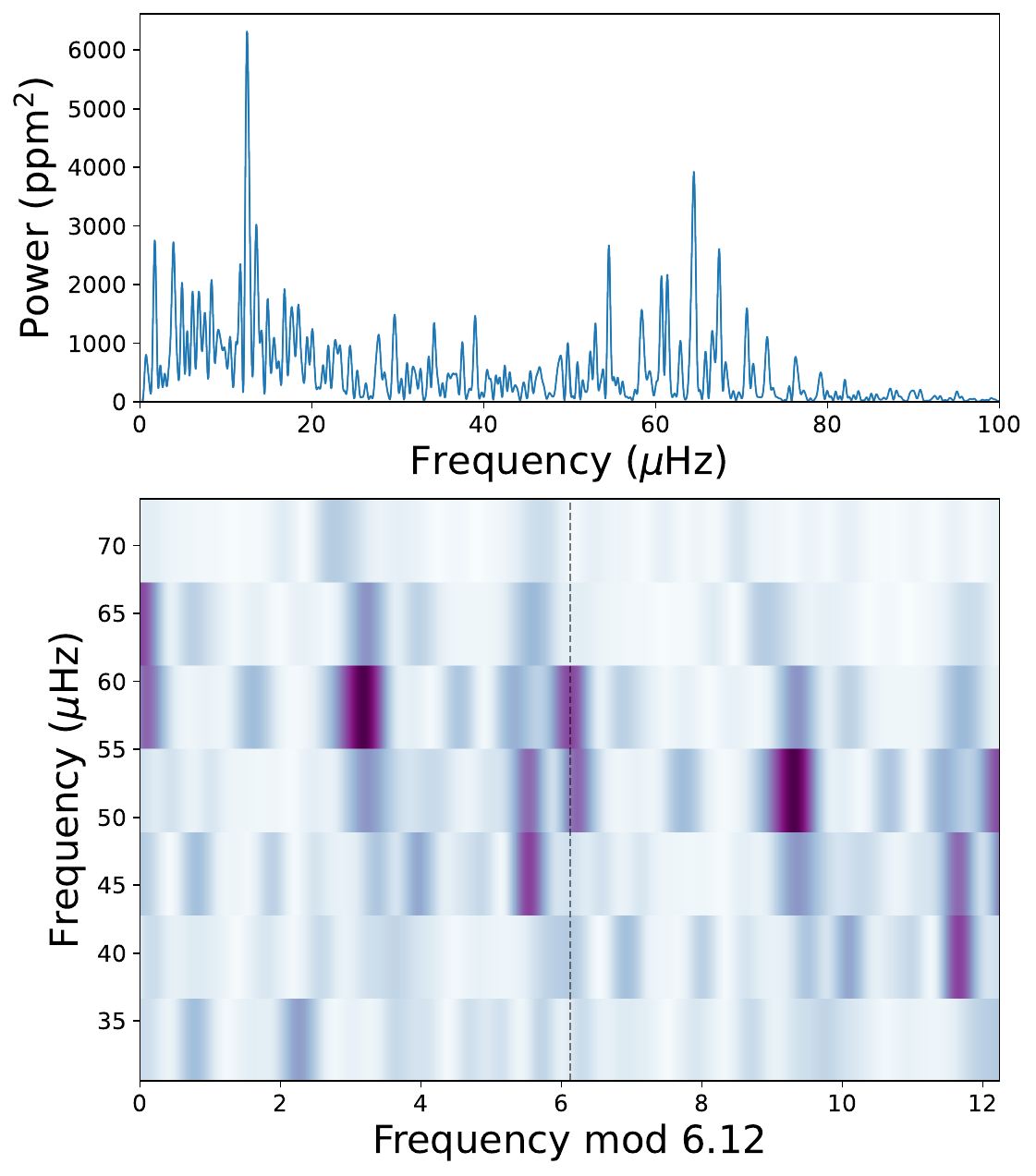}

     \caption{Upper: GLS Power spectrum of the TESS light curve of HD\,115066, showing solar-like oscillations centered at 62\,\muHz. Lower:  Power spectrum of the \textit{TESS} light curve of HD\,115066 in \echelle\ format (see text).
    } 
    \label{fig:power-spectrum-hd115066} 
\end{figure}

Oscillations in this star were measured by \citet{Hon++2021}, who reported $\numax = 66.4 \pm 3.7$\,\muHz\ based on TESS Sector~10.
We used all currently available TESS data, which comprises Sectors~10 (QLP), 37 (QLP) and 64 (TESS-SPOC), shown in Figure~\ref{fig:HD115066_lightcurve}.
The resulting power spectrum and \echelle\ diagram are shown in Fig.~\ref{fig:power-spectrum-hd115066}, with a clear detection of oscillations. We used the SYD pipeline \citep{Huber++2009} to measure $\numax = 62.2 \pm 0.8$\,\muHz\ and we used the \echelle\ diagram (see Sec.~\ref{sec:seismo-HD205577}) to measure $\Dnu = 6.12 \pm 0.03$\,\muHz. To infer the stellar properties from \numax\ and \Dnu, we adopted the following properties:
$\teff = 4632 \pm 92$\,K \citep{Hon++2021};
${\rm [Fe/H]} = -0.13 \pm 0.06$  \citep{aguilera23};
$[\alpha/{\rm Fe}] = 0.0$ (in the absence of a measurement); and
$L = 29.7 \pm 1.0 \,\Lsun$ \citep{Hon++2021}.

We fitted the observables to the grid of red giant models calculated by \citet{Stello++2009}. The five observational constraints (\numax, \Dnu, \teff, $L$ and [Fe/H]) were used to independently construct the $\chi^2$ statistics. Stellar parameters were estimated by marginalizing the likelihood functions, which are proportional to $\exp(-\chi^2/2)$. The model fits gave the following stellar parameters for HD\,115066:
$R= 8.36 \pm 0.14\,\rsun$,
$M= 1.26 \pm 0.06\,\msun$ and 
${\rm age} = 4.2 \pm 0.7$\,Gyr.

\section{Orbit Fitting and Results}\label{sec:Fitting}

\subsection{Radial Velocity Fits}

\begin{figure}
    \centering
 
     \includegraphics[width=\linewidth]{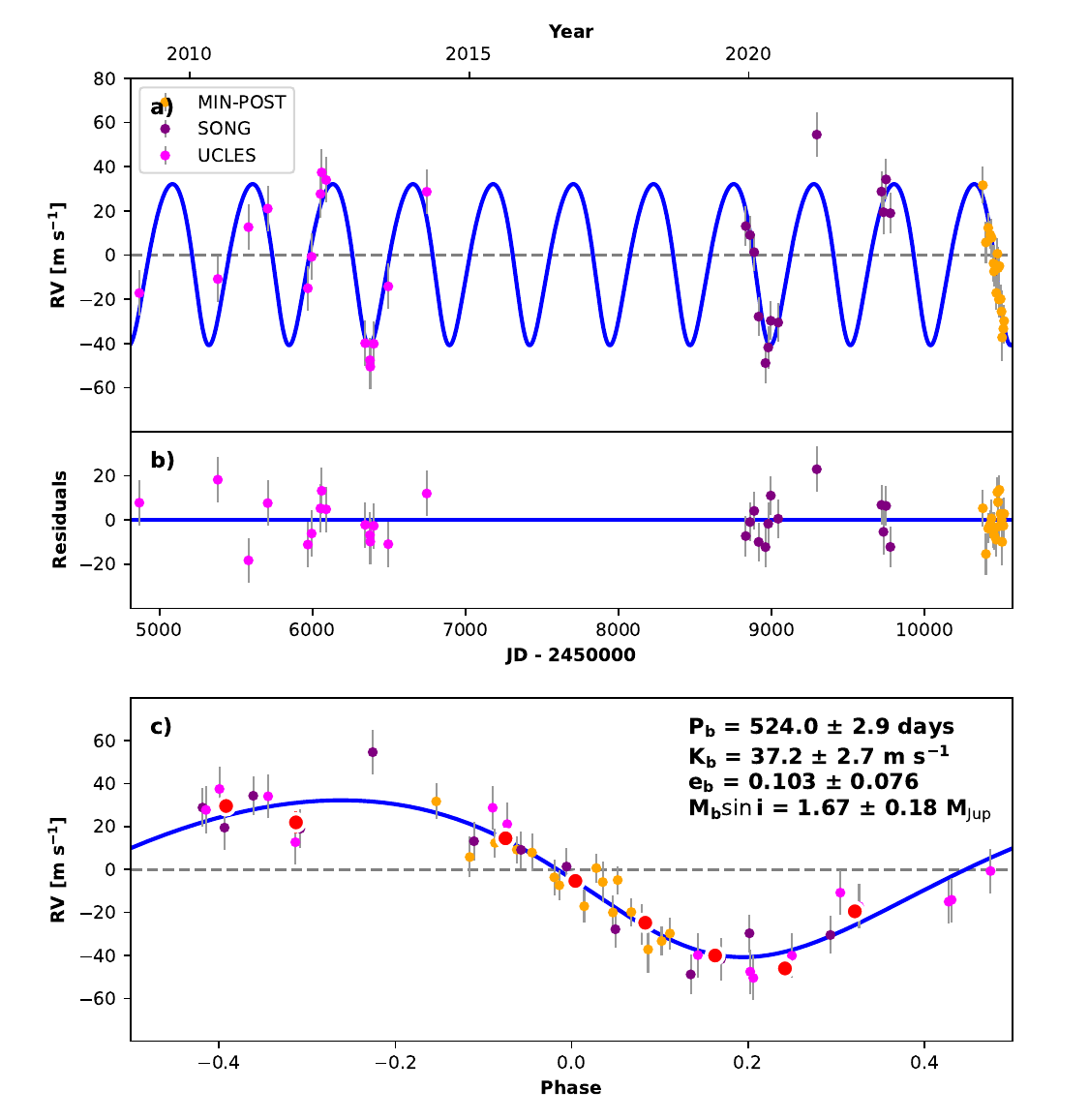}

     \caption{Radial velocity data and model fit for HD\,126105b.  The maximum likelihood model is plotted while the orbital parameters listed in Table~\ref{tab:planets} are the median values of the posterior distributions. The thin blue line is the best fit 1-planet model. b) Residuals to the best fit 1-planet model. c) RVs phase-folded to the ephemeris of planet b. The small point colors and symbols are the same as in panel a. Red points are the same velocities binned in 0.08 units of orbital phase. The phase-folded model for planet b is shown as the blue line. 
     } 
    \label{RV_plots} 
\end{figure}

For HD\,126105, we perform Keplerian orbit fits to the radial velocity data using \texttt{radvel} \citep{radvelpaper}.  Three RV data sets are included: AAT/UCLES \citep{ppps8}, SONG, and the \textsc{MINERVA}-Australis spectra that were used above for stellar parameter determination.  The one-planet fit is shown in Figure~\ref{RV_plots}.  The new SONG data were instrumental in confirming this candidate, and highlights the utility of so-called ``filler queue'' observations for this purpose.  The planet has a mass of m~sin~$i=$1.67$^{+0.19}_{-0.17}$ \mj, eccentricity $e=$0.103$^{+0.083}_{-0.069}$, and orbits at $a=$1.36$^{+0.05}_{-0.06}$ (Table~\ref{tab:planets}). 

\subsection{Astrometric Evidence} \label{subsec:astrometry}

Three of our targets -- namely HD\,115066, HD\,121156, and HD\,205577 -- show strong evidence for massive, long-period companions in our RV observations. We therefore explore the possibility of extracting further information on the nature of the companions to our targets using astrometry. All six of our target stars are present in both the \textit{Hipparcos} \citep{Hipparcos, HipparcosNew} and \textit{Gaia} \citep{Gaia, GaiaDR3}, allowing us to make use of cross-calibrated \textit{Hipparcos-Gaia} astrometry to explore long-term variations in the tangential velocities of our targets, complementing our RV observations. We query the \textit{Gaia}~DR3 version of the \textit{Hipparcos-Gaia} Catalog of Accelerations \citep[HGCA;][]{Brandt2021} for each of our targets. Four of these stars, HD\,115066 (HIP\,64647), HD\,121156 (HIP\,67890), HD\,126105 (HIP\,70382), and HD\,205577 (HIP\,106699), display significant astrometric accelerations. In the HGCA, the goodness-of-fit values for the constant proper motion hypothesis for these three giant stars are $\chi^2$ = 4659, 5416, 3068, and 309 respectively, which all lie far above the $\chi^2$ = 11.8 criterion that equates to a 0.3\% Gaussian false alarm rate \citep{Brandt2021}. In contrast, HD\,114899 and HD\,159743 have insignificant values of $\chi^2$ = 0.5 and 0.6 respectively, which accords with their absence of coherent RV variability (Section~\ref{subsec:rejected}).

For HD\,126105, the magnitude of the astrometric variability is somewhat surprising given the comparatively small RV signal. The RV solution described in the preceding section has a semi-amplitude $K$ of 37.2$^{+2.8}_{-2.6}$~\ms{}, whereas in the HGCA the net difference between the \textit{Gaia} and time-averaged \textit{Hipparcos-Gaia} proper motion is $\sim$2.19~\masyr{}, equivalent to a net change in tangential velocity of $\Delta$v $\approx$ 850~\ms{} at the distance of the star. If the astrometric signal was generated by HD\,126105\,b, this might imply that the true mass might be much larger than the RV minimum mass.

However, several lines of evidence suggest that the astrometric acceleration has a separate origin. The \textit{Gaia}~DR3-based \textit{Hipparcos-Gaia} astrometry has low sensitivity to orbits with periods shorter than the 1038~day observing duration of \textit{Gaia}~DR3, and notably has asymptotes in sensitivity at the harmonics of this duration. The 524$\pm$2.9~d orbital period of HD\,126105\,b lies at almost exactly the first harmonic of the \textit{Gaia}~DR3 duration (i.e. 1038 / 2 = 519~days), meaning that in order to generate the observed proper motion anomaly the true mass would therefore have to be extremely large, probably in the stellar regime. In this case we would expect to observe an excess in the \textit{Gaia} Renormalised Unit Weight Error (RUWE), which is highly sensitive to binaries with similar orbital periods \citep[e.g.][]{Belokurov2020, CastroGinard2024}. However, the \textit{Gaia}~DR3 RUWE is only 1.03, where RUWE = 1 suggests an ideal fit to the astrometry \citep{GaiaEDR3astrometry}, suggesting that the astrometric motion of HD\,126105 was effectively linear during \textit{Gaia} observations. We might also expect to observe a significant difference between the \textit{Gaia}~DR2 and DR3 proper motions since these have different timespans, however the two sets of parameters are identical at the 1$\sigma$ level ($\Delta\mu_\alpha$, $\Delta\mu_\delta$ = [+0.14$\pm$0.10, +0.04$\pm$0.08]~\masyr{}). All of this suggests that the proper motion of HD\,126105 does not vary at a timescale consistent with the orbital period of HD\,126105\,b.

Instead, the astrometric acceleration of HD\,126105 appears to reflect a second long-period companion in the system, one that is not presently evident in the RVs. Given the large velocity anomaly this may be a stellar companion, reminiscent of systems such as HD\,59686 \citep{Ortiz2016, Trifonov2018.59686}. However, since we do not observe any significant RV acceleration corresponding to a second companion on a wider orbit than HD\,126105\,b, this remains hypothetical. To explain the apparent absence of a significant RV acceleration it is for instance possible that the RVs may cover an inopportune orbital phase, or that the orbit for the hypothetical second companion may be observed near to pole-on.

In contrast, the three remaining targets, HD\,115066 (HIP\,64647), HD\,121156 (HIP\,67890), and HD\,205577 (HIP\,106699), simultaneously demonstrate clear evidence for massive long-period companions in both the RVs and astrometry. We therefore attribute the highly significant astrometric accelerations to the same massive long-period companions evident in the RVs, and perform joint fits to both RVs and astrometry.

\subsection{Radial Velocity + Astrometric Joint Fits} \label{subsec:joint_model}

For HD\,115066, HD\,121156, and HD\,205577, we simultaneously model our RV data with the available \textit{Hipparcos-Gaia} astrometry. The addition of astrometric information allows us to constrain the orbital inclinations ($i$) for their companions, hence resolving the RV $\sin i$ degeneracy and providing their true masses. For this purpose we use a model based on the one developed for \citet{Venner2021} with minimal modifications. Briefly, this model allows us to jointly fit the RVs and \textit{Hipparcos-Gaia} astrometry for each system using a two-body Keplerian model, with the following variable parameters: the parallax ($\varpi$) and stellar mass ($M_\odot$) for the visible primary star in each system, with Gaussian priors derived from Table~\ref{tab:allstars}; seven variables to describe the companion and its orbit, namely the true companion mass ($m$), orbital period ($P$), eccentricity and argument of periastron ($e$ and $\omega$, reparameterized as $\sqrt{e}\sin\omega$ and $\sqrt{e}\cos\omega$), the epoch of periastron ($T_\text{P}$), the orbital inclination ($i$), and the longitude of ascending node ($\Omega$); two proper motion parameters corresponding to the constant motion of the system barycenter ($\mu_{\alpha,\text{bary}}$, $\mu_{\delta,\text{bary}}$); and finally an offset and jitter parameter for each RV dataset for each relevant star. All parameters are sampled uniformly within the physical parameter space except for $P$ and $m$, which are sampled log-uniformly, and $i$, which is sampled uniformly in $\sin i$.

Since the proper motion data used in this model are time-averaged over the observing baselines of both \textit{Hipparcos} and \textit{Gaia}, we use the \texttt{htof} package \citep{htof} to construct the corresponding observing cadences (actual in case of \textit{Hipparcos}; simulated in case of \textit{Gaia}), and resample the model over the resulting set of epochs.

\begin{figure*}
    \centering
    \includegraphics[width=\columnwidth]{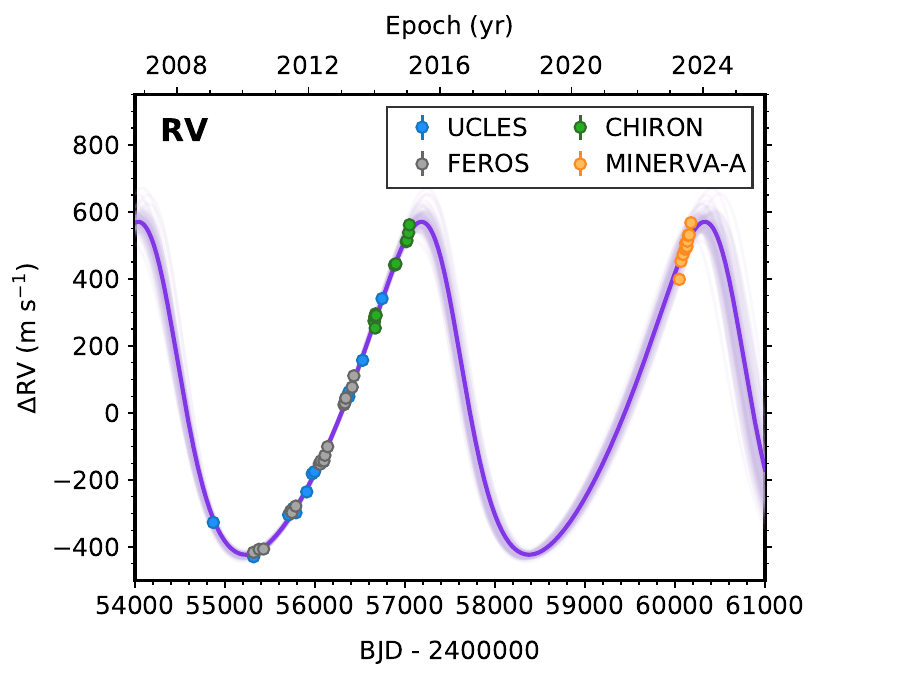}
    \includegraphics[width=\columnwidth]{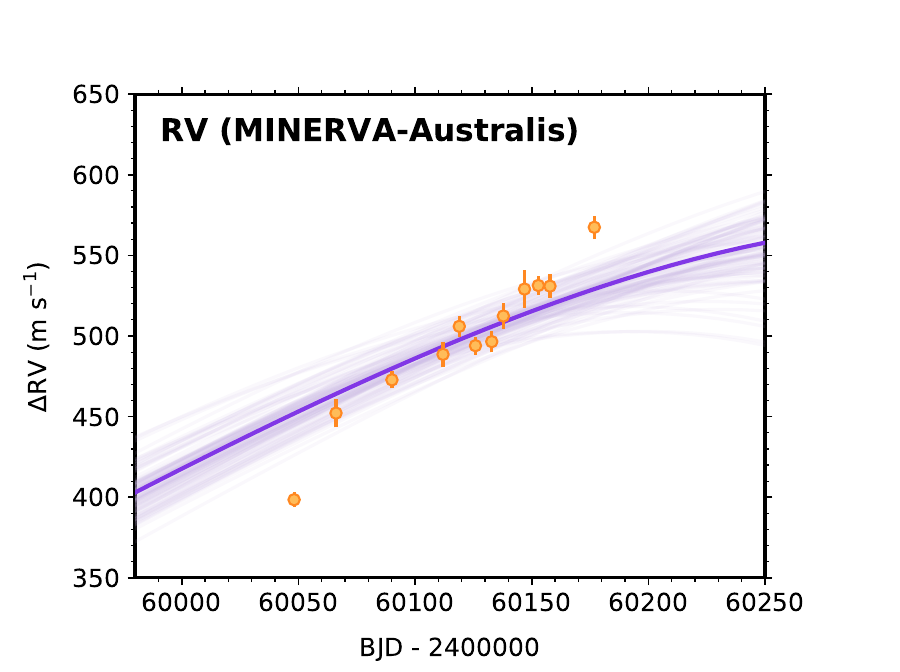}
    \includegraphics[width=\columnwidth]{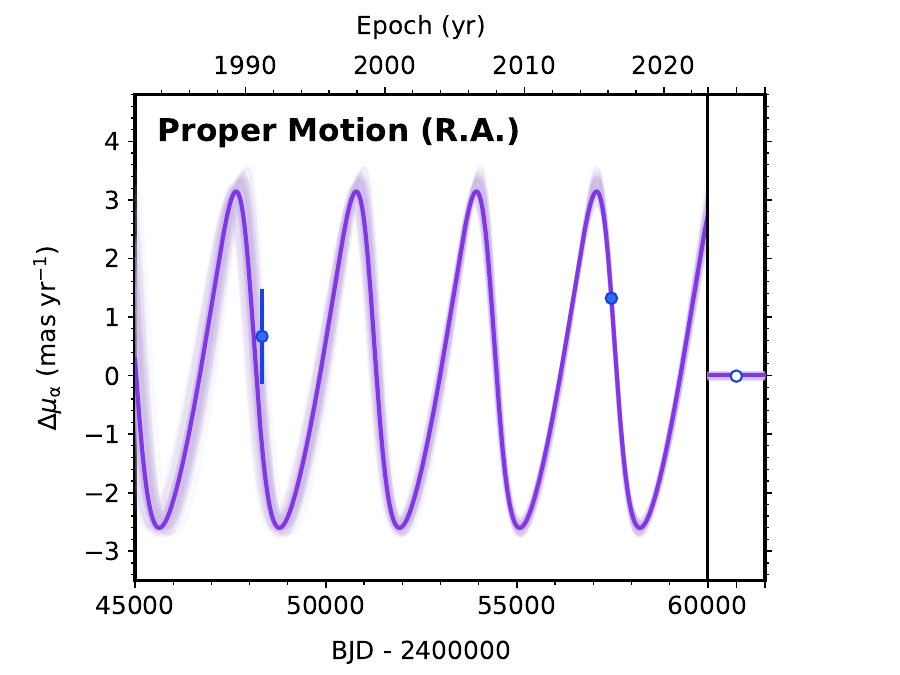}
    \includegraphics[width=\columnwidth]{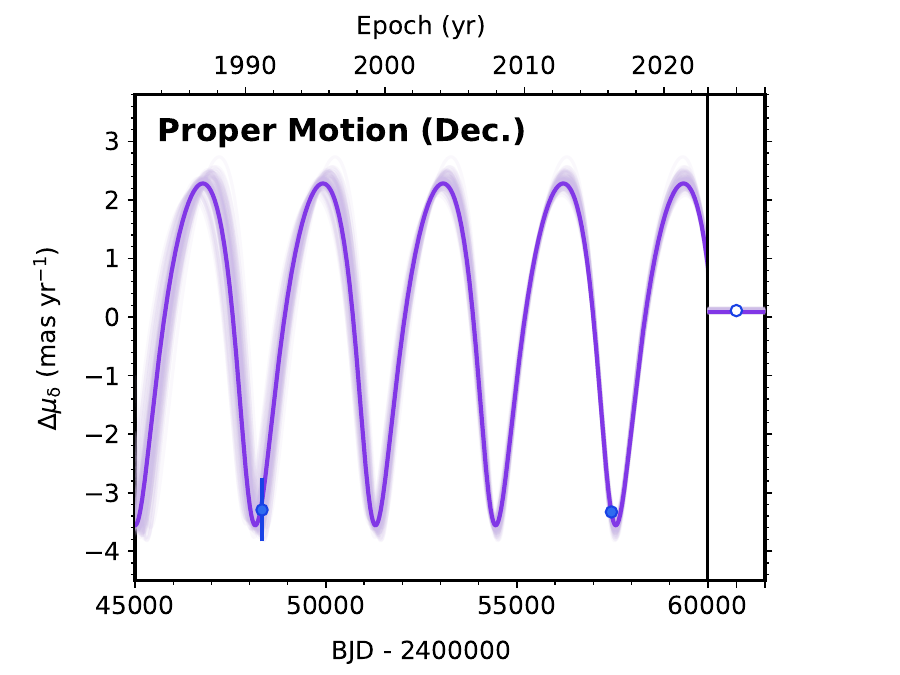}
    \caption{Joint RV and \textit{Hipparcos-Gaia} astrometry model for HD\,115066. (Top left) all RV data and best-fit RV model. (Top right) RV inset focusing on our new \textsc{Minerva}-Australis observations. (Bottom left) proper motion in right ascension and best-fit astrometric model. (Bottom right) proper motion in declination and best-fit astrometric model. In each panel the dark lines represent the best-fit model, while the light lines are drawn randomly from the posteriors. The joint model presents a well-constrained orbit for HD\,115066\,B with $P$ = 3137$^{+44}_{-39}$~days. However, while the RVs suggest a substellar minimum mass of 47.5$^{+2.7}_{-2.6}$~\mj{}, the astrometry imply a low orbital inclination ($i$ = 11.0$^{+0.5}_{-0.4}$ degrees) that results in a stellar true mass of 250$^{+16}_{-15}$~\mj{}.
    }
    \label{fig:HD115066}
\end{figure*}

While this approach worked well for HD\,115066 and HD\,121156, limitations and incompleteness of the observational data for HD\,205577 meant that the same model failed to achieve a unique orbital solution for its companion. As a result, we attempted to incorporate additional information from the \textit{Gaia} RVs,  calibrating our \textsc{Minerva}-Australis observations into the same RV system. We detail the more involved analysis developed for this system and the corresponding parameters derived for the companion in Section~\ref{subsec:HD205577}.

\subsubsection{HD 115066}

HD\,115066 (HIP\,64647) was initially reported as a spectroscopic binary with undetermined orbital parameters in \citet{bluhm16}, based on RV data from the UCLES, FEROS, and CHIRON spectrographs collected through the PPS and EXPRESS RV surveys. Based on the same UCLES dataset, \citet{ppps8} reported the first orbital solution for HD\,115066\,B with $P$ = 2817$\pm$140~days, $e$ = 0.31$^{+0.06}_{-0.05}$, and $m\sin i$ = 35$\pm$7~\mj{}. This value for the minimum mass notionally makes HD\,115066\,B a candidate brown dwarf; however, its true nature depends on the orbital inclination $i$, which cannot be constrained with RV data alone.

We acquired 12 new RV observations of HD\,115066 with \textsc{Minerva}-Australis in 2023. The first of these new observations was 2999~days subsequent to the most recent prior observation, approximately equivalent to one complete orbital revolution of HD\,115066\,B. The \textsc{Minerva}-Australis observations appear to fall near to the RV maximum of the companion's orbit.

In the HGCA, the net difference between the \textit{Gaia} and time-averaged \textit{Hipparcos-Gaia} proper motion is a large $\sim$2.56~\masyr{}, equivalent to a net change in tangential velocity of $\Delta$v $\approx$ 2300~\ms{} at the distance of HD\,115066. The variation in tangential velocity is significantly higher than the $\approx$1000~\ms{} peak-to-peak variation observed in RV, which suggests that $\sin i$ is likely to be relatively low and that the true mass of HD\,115066\,B is larger than the RV minimum mass (compare \citealt{Venner2021} for a similar scenario).

\begin{figure*}
    \centering
    \includegraphics[width=\columnwidth]{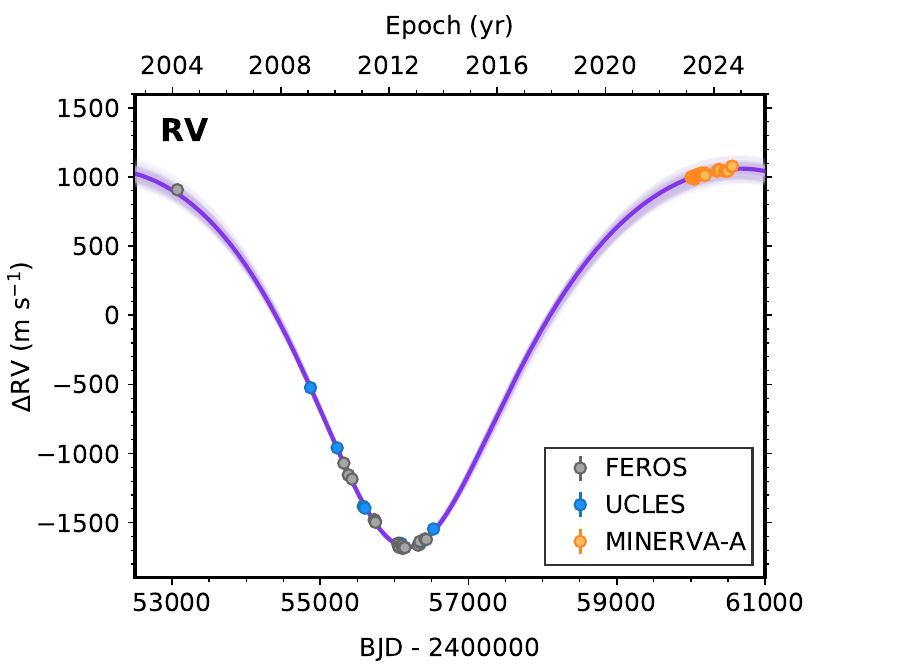}
    \includegraphics[width=\columnwidth]{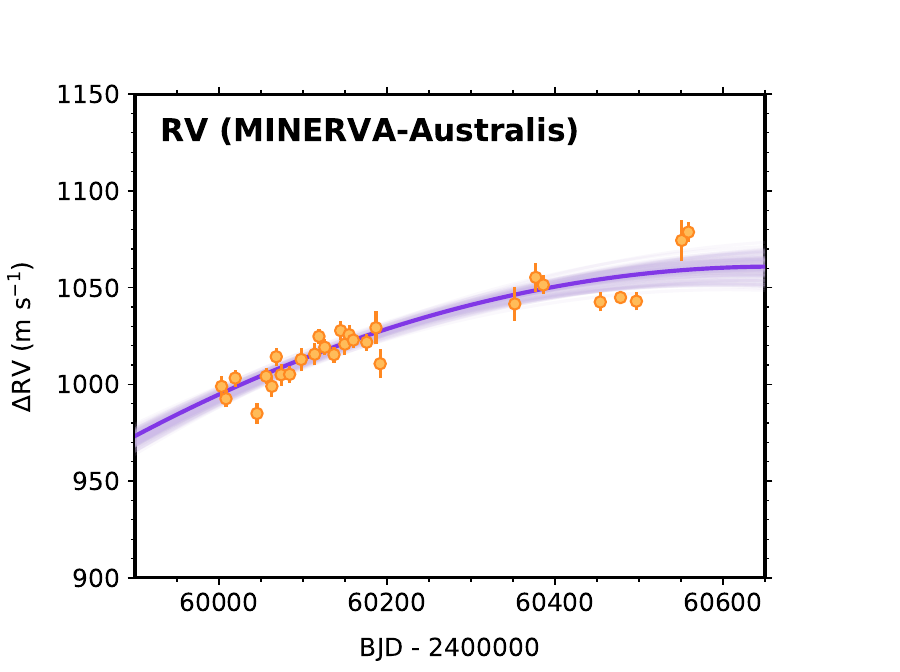}
    \includegraphics[width=\columnwidth]{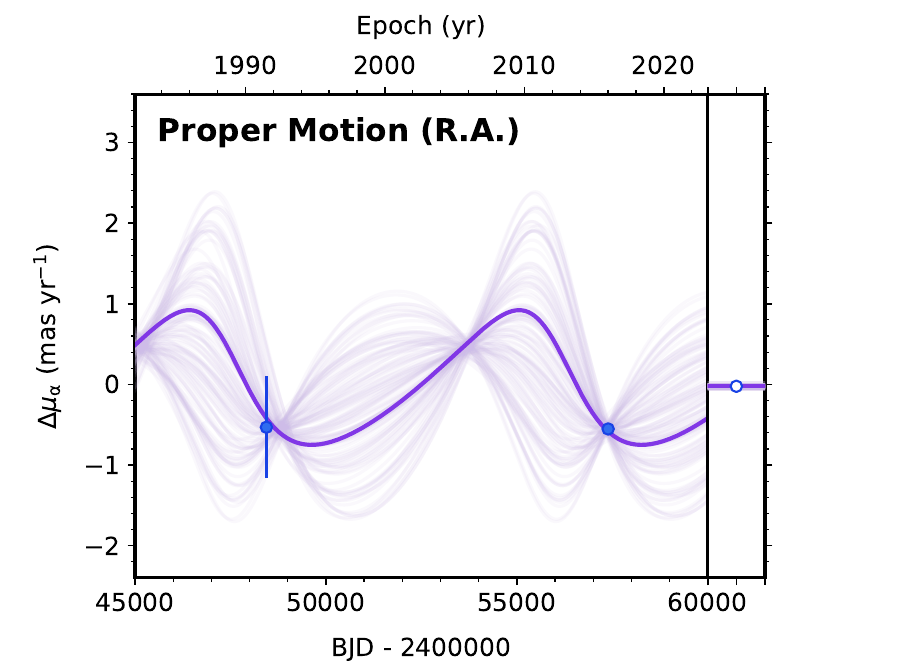}
    \includegraphics[width=\columnwidth]{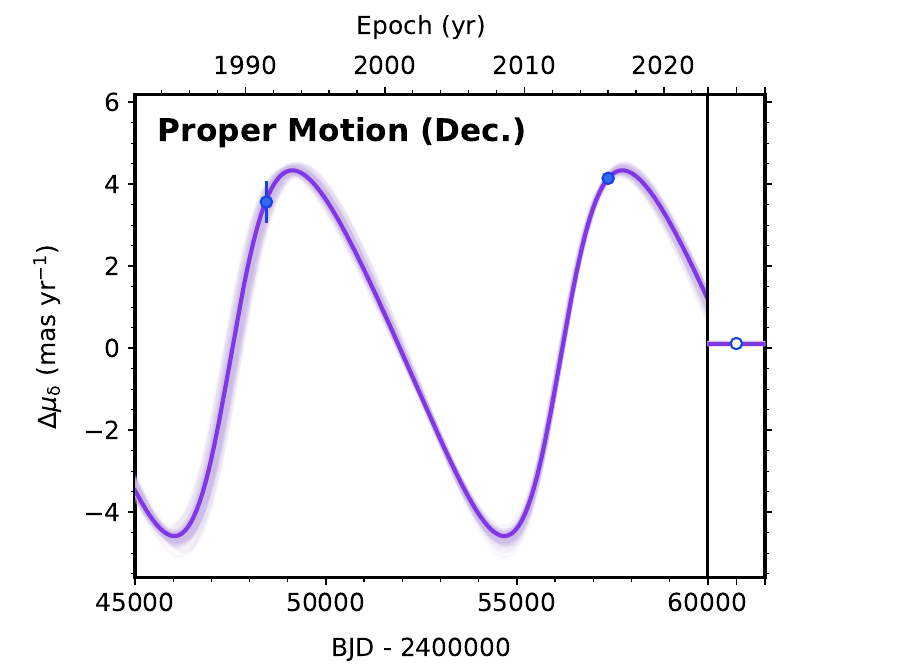}
    \caption{Joint RV and \textit{Hipparcos-Gaia} astrometry model for HD\,121156. (Top left) all RV data and best-fit RV model. (Top right) RV inset focusing on our new \textsc{Minerva}-Australis observations. (Bottom left) proper motion in right ascension and best-fit astrometric model. (Bottom right) proper motion in declination and best-fit astrometric model. In each panel the dark lines represent the best-fit model, while the light lines are drawn randomly from the posteriors. The joint model returns an orbital solution for HD\,121156\,B consistent with edge-on ($i$ = 90$\pm$9~degrees) with $P$ = 8604$^{+95}_{-89}$~days and a stellar true mass of 179$\pm$5~\mj{}.
    }
    \label{fig:HD121156}
\end{figure*}

We have performed a joint fit to the RVs and \textit{Hipparcos-Gaia} astrometry, making use of all available UCLES, FEROS, CHIRON, and \textsc{Minerva}-Australis RV data. We present the results of this joint model in Figure~\ref{fig:HD115066}, and report the corresponding posterior parameters in Table~\ref{tab:planets}.  Our solution for HD\,115066\,B is similar to but significantly more precise than the parameters reported in \citet{ppps8}, with $P$ = 3137$^{+44}_{-39}$~days, $e$ = 0.249$^{+0.014}_{-0.013}$, and $m\sin i$ = 47.5$^{+2.7}_{-2.6}$~\mj{}. The RV acceleration visible in the \textsc{Minerva}-Australis observations appears at first sight to be larger than expected from the model; however, this can be explained by the first and last observations being few-$\sigma$ outliers, the first having a negative displacement and the last being positively displaced. The remaining 10 data points agree well with the model. 

Unfortunately, as suspected based on our provisional evaluation of the \textit{Hipparcos-Gaia} astrometry, the joint model finds that the orbit of the companion is viewed near to pole-on with $i$ = 11.0$^{+0.5}_{-0.4}$ degrees ($\sin i$ = 0.190$\pm0.008$), resulting in a true mass for HD\,115066\,B of $m$ = 250$^{+16}_{-15}$\,\mj{} (0.238$^{+0.015}_{-0.014}$~$M_\odot$) that is well above the stellar-substellar transition. Hence the addition of astrometry reveals that the companion of HD\,115066 is a star, possibly an M-dwarf, instead of a brown dwarf.

\subsubsection{HD 121156}

As with the previous example of HD\,115066, HD\,121156 (HIP\,67890) was also first reported as a spectroscopic binary in \citet{bluhm16} with an incomplete orbit, based on RVs collected with the UCLES and FEROS spectrographs. Based solely on the UCLES data, \citet{ppps8} reported an orbital solution with $P$ = 3033$^{+470}_{-420}$~days, $e$ = 0.13$^{+0.07}_{-0.05}$, and $m\sin i$ = 54$\pm$11~\mj{}. Thus HD\,121156\,B could in principle be a brown dwarf. However, one suspects that the orbital period may be underestimated; the FEROS data reported in \citet{bluhm16} approximately doubles the total observational baseline, to approximately $\approx$3000~days, but it is not consistent with a $\approx$3000 day orbital period. If $P$ is significantly larger than found in \citet{ppps8}, then it is likely that $m\sin i$ will be larger as well.

We acquired 29 new RVs of HD\,121156 across the 2023-2024 observing seasons with \textsc{Minerva}-Australis. The first of these was 3475 days later than the most recent UCLES observation of the star. The new \textsc{Minerva}-Australis observations capture a quadratic acceleration prior to RV maximum.

In the HGCA, the net change in proper motion is $\sim$3.94~\masyr{}, equivalent to a change in tangential velocity of $\Delta$v $\approx$ 1200~\ms{} given the distance to HD\,121156. This is larger than the RV semi-amplitude reported by \citet{ppps8}, but appears more compatible with the $\approx$2600~\ms{} peak-to-peak variation apparent in \citet{bluhm16}. The proper motion variability is disproportionately seen in the declination component ($\Delta\mu_\alpha$ = -0.51~\masyr{}, $\Delta\mu_\delta$ = +3.90~\masyr{}), suggesting that the orbit of HD\,121156\,B is likely to be primarily observed projected north-south.

\begin{figure*}
    \centering
    \includegraphics[width=\columnwidth]{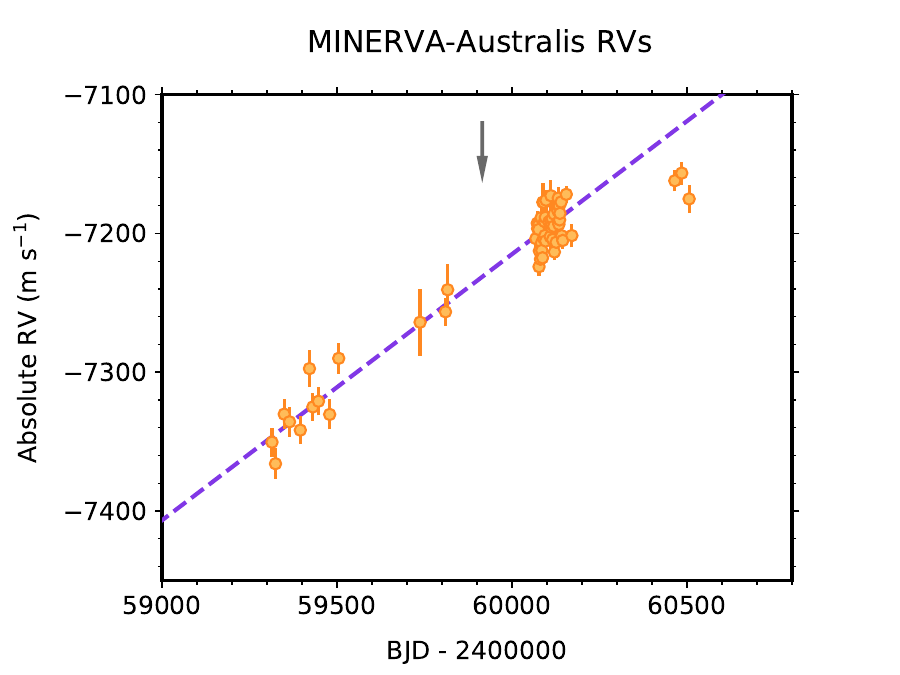}
    \includegraphics[width=\columnwidth]{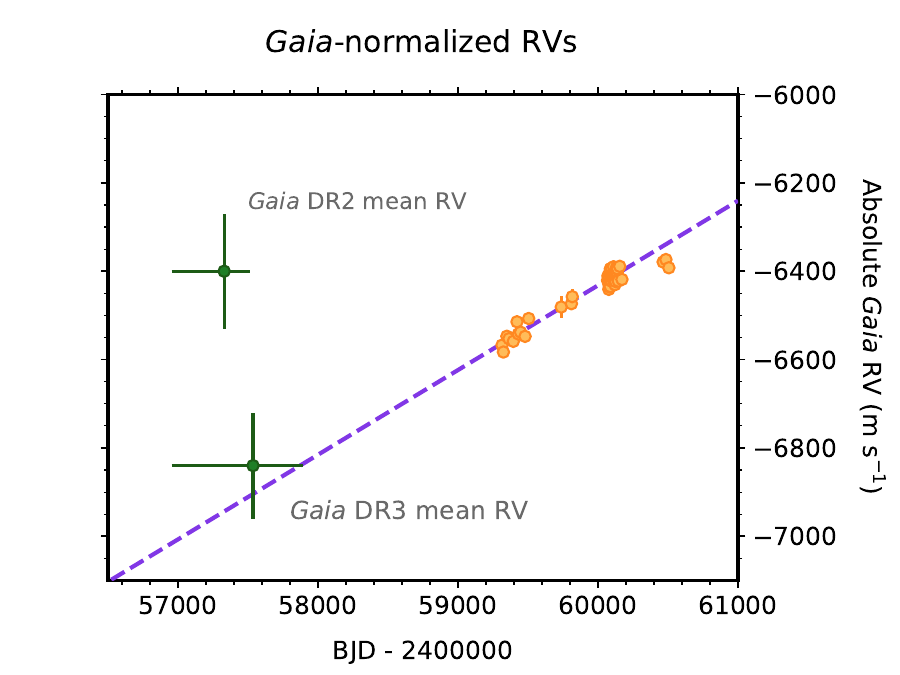}
    \caption{Absolute RV data for HD\,205577. (\textit{Left}) Four seasons of \textsc{Minerva}-Australis observations, with a fiducial 70~\msyr{} acceleration shown as a visual guide. The epoch of the instrument intervention is identified with an arrow. (\textit{Right}) As before, but shifted to the \textit{Gaia} RV reference frame and with the addition of mean RVs from \textit{Gaia}~DR2 and DR3. While the RV observations are reasonably consistent with a linear acceleration, the \textit{Gaia}~DR2 mean RV and the 2024 \textsc{Minerva}-Australis observations appear to diverge, suggesting orbital curvature.}
    \label{fig:HD205577_RV}
\end{figure*}

The result of our joint fit combining the UCLES, FEROS, and \textsc{Minerva}-Australis RVs with \textit{Hipparcos-Gaia} astrometry are displayed in Figure~\ref{fig:HD121156}, and the companion parameters are tabulated in Table~\ref{tab:planets}. This solution differs significantly from the one reported in \citet{ppps8} primarily in that HD\,121156\,B has a much longer orbital period, $P$ = 8604$^{+95}_{-89}$~days (23.6~years). This is coincidentally similar to the 25-year interval between the observations of \textit{Hipparcos} and \textit{Gaia}, hence the time-resolved proper motions sample almost identical orbital phases. The time-averaged \textit{Hipparcos-Gaia} mean proper motion, which covers the interval between \textit{Hipparcos} and \textit{Gaia} observations, provides much of the ultimate constraint on the amplitude of the peak-to-peak variation in proper motion. From our joint model we find that the orbital inclination is robustly constrained to be edge-on with $i$ = 90$\pm$9~degrees ($\sin i$ =  0.995$^{+0.005}_{-0.016}$); the minimum mass and true mass are correspondingly nearly identical, with $m\sin i$ = 177$\pm$5~\mj{} and $m$ = 179$\pm$5~\mj{} (0.171$\pm$0.005~$M_\odot$). Hence HD\,121156\,B is likewise a low-mass star instead of a brown dwarf. Other notable properties of our orbital solution are a modest orbital eccentricity (0.226$\pm$0.010) and a longitude of ascending node of $\Omega$ = 172$\pm5$~degrees, which is indeed near to north-south alignment (i.e. $\Omega$ $\equiv$ 180~degrees).

\subsubsection{HD\,205577} \label{subsec:HD205577}

\citet{Wittenmyer2020} reported 8 UCLES RV observations for HD~205577 spanning 4.0~years (2009 - 2013), which demonstrate a long-term acceleration of the stellar RV with amplitude above $>$300~\ms{}. A speculative orbital solution was proposed on the basis of these observations, with $P=1686$~d, $K=613$~\ms{}, and $e=0.97$. However, this would entail an orbital period longer than the RV duration and a rapid periastron passage lying outside of existing observations, and hence requires validation.

We have acquired 61 new RV observations with \textsc{Minerva}-Australis spanning 3.3~years (2021 - 2024) with a median uncertainty of 8.5~\ms{}, continuing on approximately 8 years after the last UCLES observation. As with the UCLES data, the RV variability observed by \textsc{Minerva}-Australis can mainly be described by a long-term acceleration. The phase coverage for the companion's orbit is therefore evidently incomplete, and the same joint RV + astrometry model used in the preceding sections fails to converge given the available data. This necessitates a more involved analysis to retrieve the properties of the companion in the system.

We begin by first placing our new \textsc{Minerva}-Australis RVs on a consistent absolute reference frame. \textsc{Minerva}-Australis underwent an instrument intervention in December 2022, which introduced an offset in the spectrograph RV zero-point. We use observations of the classical RV standard HD~10700 ($\tau$~Ceti) to calibrate for the effects of this offset. Comparing the medians of the pre- and post-intervention RVs, we calculate a  $(\text{\textsc{Minerva}-A}_{\text{post}}-\text{\textsc{Minerva}-A}_{\text{pre}})$ RV offset of $-47\pm9$~\ms{} for HD~10700.

In the left panel of Figure~\ref{fig:HD205577_RV} we show the \textsc{Minerva}-Australis RVs, normalized to the post-intervention RV system. It can be seen that the first three seasons of observations are consistent with a linear acceleration that crosses the instrument intervention, visualised with a fiducial 70~\msyr{} acceleration determined from a least-squares fit to these three seasons of RVs. However, the most recent data from the 2024 season diverge from a purely linear solution. While the magnitude of change is modest, this indicates that \textsc{Minerva}-Australis observations have captured a non-negligible fraction of the orbit of the companion.

\begin{figure*}
    \centering
    \includegraphics[width=\textwidth]{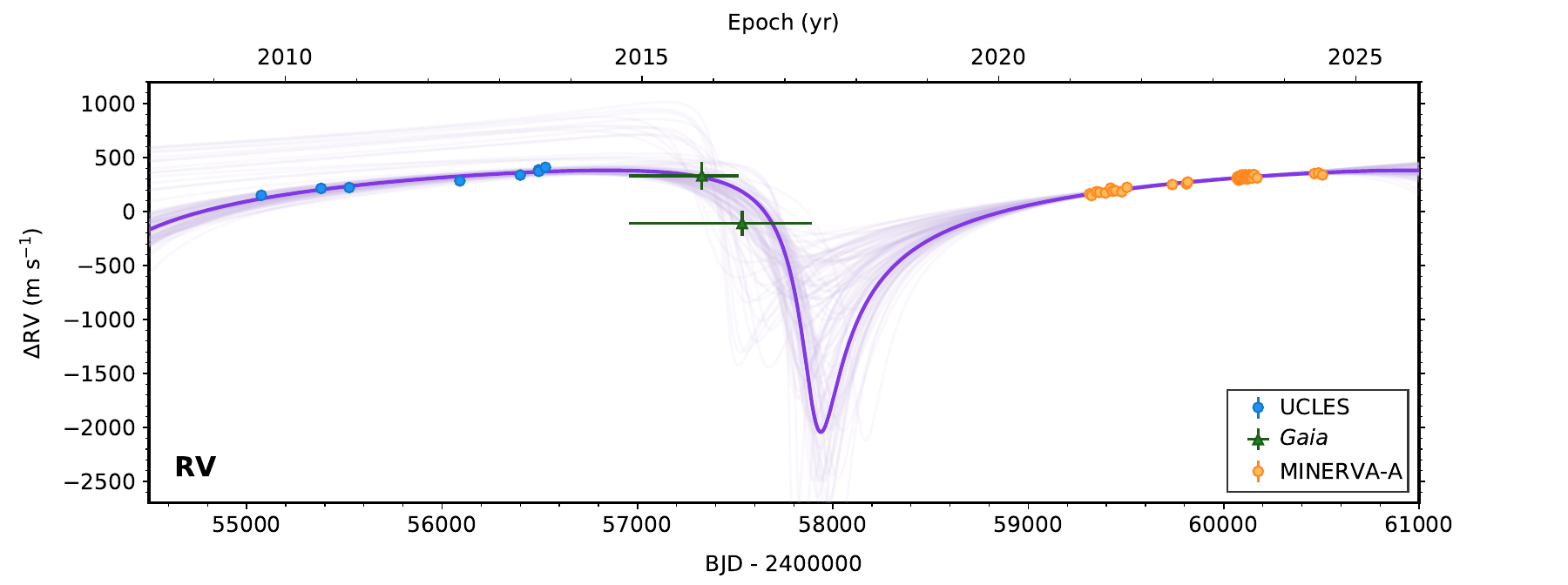}
    \includegraphics[width=\columnwidth]{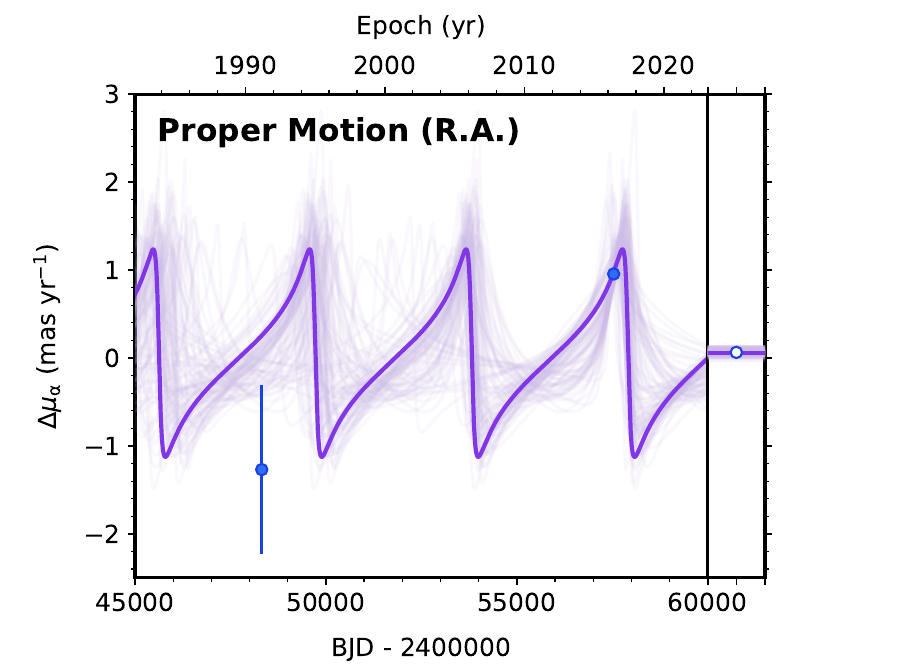}
    \includegraphics[width=\columnwidth]{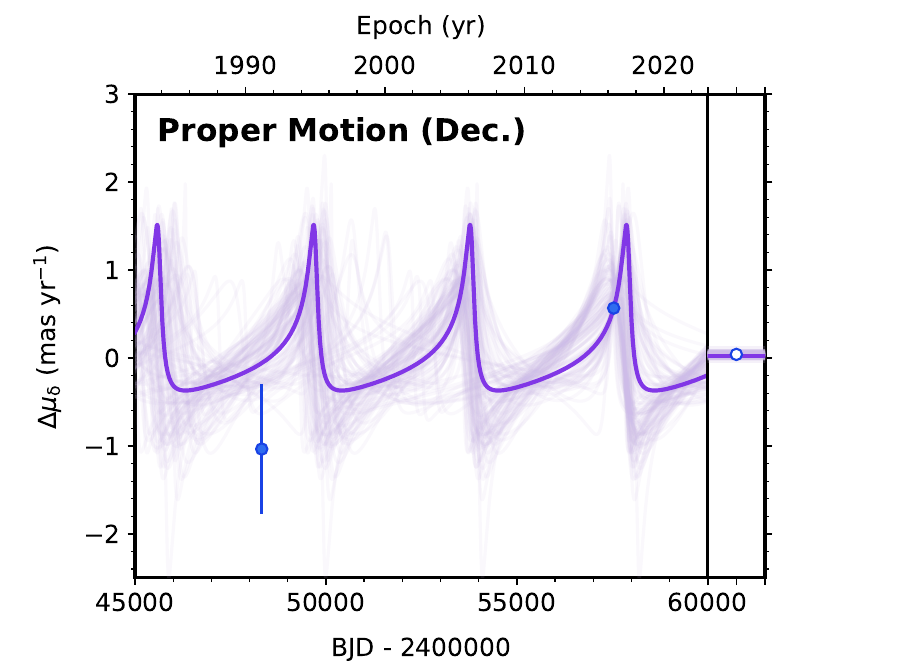}
    \caption{Joint RV and \textit{Hipparcos-Gaia} astrometry model for HD\,205577. (Top) all RV data, including \textit{Gaia} mean RVs, and best-fit RV model. (Bottom left) proper motion in right ascension and best-fit astrometric model. (Bottom right) proper motion in declination and best-fit astrometric model. In each panel the dark lines represent the best-fit model, while the light lines are drawn randomly from the posteriors. Though the orbit of HD\,205577\,B is not well-constrained by the data, the joint model returns a unique solution with $P$ = 4100$^{+800}_{-180}$~days, a high eccentricity of 0.68$^{+0.10}_{-0.16}$, and a true mass for the companion of 77$^{+11}_{-9}$~\mj{} that straddles the Hydrogen-burning limit.
    }
    \label{fig:HD205577_model}
\end{figure*}

As UCLES RVs are relative (instead of absolute), we cannot directly place these into the same reference frame as the \textsc{Minerva}-Australis RVs. However, we can marshal absolute RVs measured by \textit{Gaia} to greatly extend our absolute RV baseline. both \textit{Gaia}~DR2 and \textit{Gaia}~DR3 have reported time-averaged mean RVs measured by the \textit{Gaia} Radial Velocity Spectrometer. This allows us to extend the absolute RV baseline backwards to $\sim$2015, facilitating a stronger comparison with the preceding UCLES RVs. To do this, we must first convert between the \textit{Gaia} and \textsc{Minerva}-Australis RV reference frames. For this purpose we again turn to observations of HD~10700. The median post-intervention \textsc{Minerva}-Australis RV of HD~10700 is $-17.380\pm0.008$~\kms{}, which compares to $-16.597\pm0.0002$~\kms{} in the \textit{Gaia} reference frame \citep{Soubiran2018}.\footnote{The \textit{Gaia} Radial Velocity Spectrometer is not internally calibrated. \citet{Soubiran2018} provide the main part of the zero-point calibration using observations from ground-based spectrographs. In effect, the cited value is the ``underlying" RV in the \textit{Gaia} system.} Hence the $(\text{\textsc{Minerva}-A}-\textit{Gaia})$ systematic RV offset is calibrated to -0.783$\pm$0.008~\kms{}.

We may validate this approach by considering the \textsc{Minerva}-Australis observations of HD~126105. In our orbital solution, the \textsc{Minerva}-Australis RV system offset is $42.223\pm0.007$~\kms{}. Applying the RV offset of $-0.783$~\kms{}, this results in a system RV of $43.01$~\kms{} in the \textit{Gaia} reference frame. The actual mean RV in \textit{Gaia}~DR3 is in fact $43.05\pm0.12$~\kms{} \citep{GaiaDR3}. At this level of RV precision, we can ignore the orbital motion from the planet for this comparison. We therefore conclude that there is agreement between the \textit{Gaia} and \textsc{Minerva}-Australis RV zero-points after this empirical recalibration at the $<$0.1~k\ms{} level.

In the right panel of Figure~\ref{fig:HD205577_RV} we show the \textit{Gaia}~DR2 and DR3 mean RVs placed on the same scale as the \textsc{Minerva}-Australis RVs, extrapolating the previous linear acceleration model backwards to the \textit{Gaia} epochs. While the \textit{Gaia}~DR3 mean RV is congruent with this simple model, the \textit{Gaia}~DR2 mean RV is $\approx$3$\sigma$ higher than both the DR3 measurement and the model extrapolation. As the DR2 and DR3 RVs are calibrated in a fundamentally similar way \citep{Katz2019, Katz2023}, we see it as unlikely that this effect is an instrument effect; instead, it appears more plausible to assume that the RV of HD~205577 genuinely decreased during the course of \textit{Gaia} RV observations, which becomes evident in the mean RVs due to the increased time baseline of \textit{Gaia}~DR3.

In the HGCA, the net difference between the \textit{Gaia} and mean \textit{Hipparcos-Gaia} proper motions is $\sim$$1.08$~\masyr{}, equivalent to a tangential velocity change $\Delta v \approx1100$~\ms{} at the distance of the star. The observed proper motion variability is highly significant \citep[$\chi^2=309$,][]{Brandt2021}, and points to a variation in tangential velocity larger than what has been captured by our RV observations.

Having now assembled the available evidence for the orbital motion of HD\,205577, we finally attempt to identify a satisfactory orbital solution. Here we assume the existence of a single companion with a Keplerian orbit. Our joint model to the RVs and \textit{Hipparcos-Gaia} astrometry is based on the one from \citet{Venner2021}, as described in Section~\ref{subsec:joint_model}, with modifications to accommodate for the \textit{Gaia} RV data. To do this, we use the \textsc{Minerva}-Australis RVs renormalized to the \textit{Gaia} RV zero-point, and then introduce the \textit{Gaia}~DR2 and DR3 mean RVs to the model using the same RV offset parameter.

Since the \textit{Gaia} RVs are time-averaged, like the proper motions used in the astrometry model, we must resample the model using the underlying observing epochs; to estimate these epochs, we again use \texttt{htof} \citep{htof} to reconstruct the \textit{Gaia} observing cadence for HD\,205577. When consecutive \textit{Gaia} epochs are spaced within $<$0.5~days, we take the medians of these epochs so as to down-weight times where \textit{Gaia} observed the star several times in rapid succession. We then institute time cuts appropriate for the observing durations of \textit{Gaia}~DR2 and \textit{Gaia}~DR3.\footnote{i.e. $[2014.562, 2016.391]$ for \textit{Gaia}~DR2 \citep{GaiaDR2} and $[2014.562, 2017.404]$ for \textit{Gaia}~DR3 \citep{GaiaEDR3, GaiaDR3}.} In this way, we estimate mean \textit{Gaia} RV epochs of BJD 2457241 and 2457483 for \textit{Gaia}~DR2 and DR3, with respective total ranges of [2456954, 2457517] and [2456954, 2457893]. We use these values for the purposes of visualization in Figure~\ref{fig:HD205577_RV} and Figure~\ref{fig:HD205577_model}.

The apparently large change in velocity observed by \textit{Gaia}, seen in both RV and astrometry and discussed in the preceding text, appears most congruent with a system architecture involving a massive companion that went through periastron passage during the \textit{Gaia} observations \citep[compare][]{Venner2022}. Combined with the picture given by our precise RV observations, this can be reconciled with a high-eccentricity orbit that has gone through approximately one whole revolution over the span of RV observations (i.e. $P\approx10$~yr).

The results of our joint model are presented in Figure~\ref{fig:HD205577_model} and Table~\ref{tab:planets}. Though the posterior constraints for HD\,205577\,B are comparatively less precise than the other systems studied in this work, we are able to identify a unique solution for the companion parameters. We find that HD\,205577\,B has an orbital period of $P$ = 4100$^{+800}_{-180}$~days and a high eccentricity of 0.68$^{+0.10}_{-0.16}$. The orbital inclination is only broadly constrained, $i$ = 90$\pm$50~degrees; however, since a significant part of the orbital constraints are provided by the astrometry, which captures a larger proportion of the net change in velocity across the orbit, the estimated true mass is in fact more precisely determined than the $m\sin i$ minimum mass, $m$ =  77$^{+11}_{-9}$~\mj{} versus $m\sin i$ = 56$^{+21}_{-19}$~\mj{}. HD\,205577\,B therefore appears to straddle the hydrogen-burning limit, the conventional distinction between stars and brown dwarfs.

Much of the parameter uncertainties in our joint model stems from poor coverage of the periastron passage, which has only been observed by the (time-averaged) \textit{Gaia} RVs and astrometry. However, our posterior constraints imply that the variation was comparatively large in both dimensions during \textit{Gaia} observations ($\approx$1~k\ms{} in RV, $\approx$1~\masyr{} in proper motion). The epoch \textit{Gaia} astrometry and RVs will both be publicly released in the future \textit{Gaia}~DR4; if our orbital solution for HD\,205577\,B is broadly correct we expect that its previous periastron passage will be detected at high significance in these datasets, which is likely to allow for significantly improved constraints on the orbital and physical parameters of the companion. On the other hand, if the \textit{Gaia}~DR4 astrometry and RVs differ in magnitude or sign of variation as compared to the predictions of our orbital model in Figure~\ref{fig:HD205577_model}, then this may point to different orbital properties for HD\,205577\,B.

\begin{table*}
	\centering
	\caption{Final parameters for the four companions confirmed in this work.}
	\label{tab:planets}
	\begin{tabular}{lrrrrr} 
		\hline
		Parameter & HD\,115066B & HD\,121156B & HD\,126105b & HD\,205577B & Units \\
		\hline
Period        &  3137$^{+44}_{-39}$  & 8604$^{+95}_{-89}$  & 524.0$\pm$2.9  &  4100$^{+800}_{-180}$   & days \\ 
$T_\text{p}$ &  2454311$\pm$46  & 2456051$^{+26}_{-29}$  &  2455291$^{+55}_{-82}$  &  2457840$^{+120}_{-230}$  &  BJD \\
\textit{K}    &  500$^{+13}_{-12}$ & 1365$\pm$20  &  37.2$^{+2.8}_{-2.6}$    &  840$^{+470}_{-350}$   &  \ms \\ 
Eccentricity  &  0.249$^{+0.014}_{-0.013}$  & 0.226$\pm$0.010  & 0.103$^{+0.083}_{-0.069}$  & 0.68$^{+0.10}_{-0.16}$  &   \\ 
$\omega$      &  50.7$^{+3.2}_{-3.0}$  & 171.7$^{+1.6}_{-1.9}$  & 155$\pm$45 &    151$^{+10}_{-21}$  &  degrees \\
\hline
$i$                & 11.0$^{+0.5}_{-0.4}$ & 90$\pm$9 &  --  &  90$\pm$50  &  degrees  \\
sin $i$            & 0.190$\pm0.008$ & 0.995$^{+0.005}_{-0.016}$ &  --  &  0.74$^{+0.22}_{-0.25}$  &    \\
$\Omega$           & 193$\pm$3 & 172$\pm5$ &  --  &  53$^{+14}_{-20}$  &  degrees  \\
$m$ sin $i$    & 47.5$^{+2.7}_{-2.6}$  & 177$\pm$5  & 1.67$^{+0.19}_{-0.17}$   &  56$^{+21}_{-19}$  &  \mj  \\
$m$            & 250$^{+16}_{-15}$ & 179$\pm$5 &  --  &  77$^{+11}_{-9}$  &  \mj  \\
$a$            & 4.89$\pm$0.13  & 9.45$\pm$0.13  &  1.36$^{+0.05}_{-0.06}$  &  5.4$^{+0.6}_{-0.3}$  &  au \\
\hline
$\gamma_\text{UCLES}$     & 172$\pm$10  & 1393$^{+19}_{-21}$ & 0.9$\pm$3.7  & -310$\pm$70   & \ms  \\
$\gamma_\text{MINERVA-A}$ & -7963$^{+35}_{-29}$  & -17594$^{+27}_{-26}$  & 222.8$^{+7.1}_{-7.4}$  & -6680$^{+90}_{-70}$$^{a}$  & \ms  \\
$\gamma_\text{SONG}$      & --  & --  & 508.0$^{+3.3}_{-3.4}$  & --  & \ms  \\
$\gamma_\text{FEROS}$     & 152$^{+11}_{-9}$  & --  & --  & --  & \ms  \\
$\gamma_\text{CHIRON}$    & -399$^{+13}_{-12}$  & 1434$^{+19}_{-21}$  & --  & --  & \ms  \\
\hline
Barycentre $\mu_{\alpha}$ & -32.68$\pm$0.03 & -168.22$\pm$0.02 & -- & -5.44$^{+0.06}_{-0.05}$ & \masyr \\
Barycentre $\mu_{\delta}$ & -17.44$\pm$0.05 & -70.09$\pm$0.04 & -- & -49.42$\pm$0.06 & \masyr \\
\hline
$\sigma_\text{UCLES}$ & 12$^{+2}_{-3}$ & 16$^{+6}_{-4}$ & 12.6$^{+3.1}_{-2.4}$  & $35^{+18}_{-10}$ &  \ms \\
$\sigma_\text{MINERVA-A}$ & 24$^{+9}_{-6}$ & 6.8$^{+1.8}_{-1.4}$ & 4.1$^{+2.8}_{-2.6}$   & 9$\pm$2 &  \ms \\
$\sigma_\text{SONG}$ & -- & --  & 10.0$^{+3.6}_{-2.7}$  & -- &  \ms \\
$\sigma_\text{FEROS}$ & 5$^{+3}_{-2}$ & 9$^{+3}_{-2}$ & -- & -- &  \ms \\
$\sigma_\text{CHIRON}$ & 14$^{+5}_{-3}$ & -- & -- & -- &  \ms \\
        \hline
	\end{tabular}
\justifying
\newline
(a) Renormalised to the \textit{Gaia} RV reference frame as per Section~\ref{subsec:HD205577}.
\end{table*}

\subsection{Two Candidates Rejected} \label{subsec:rejected}

\begin{figure}
    \centering
 
     \includegraphics[width=\linewidth]{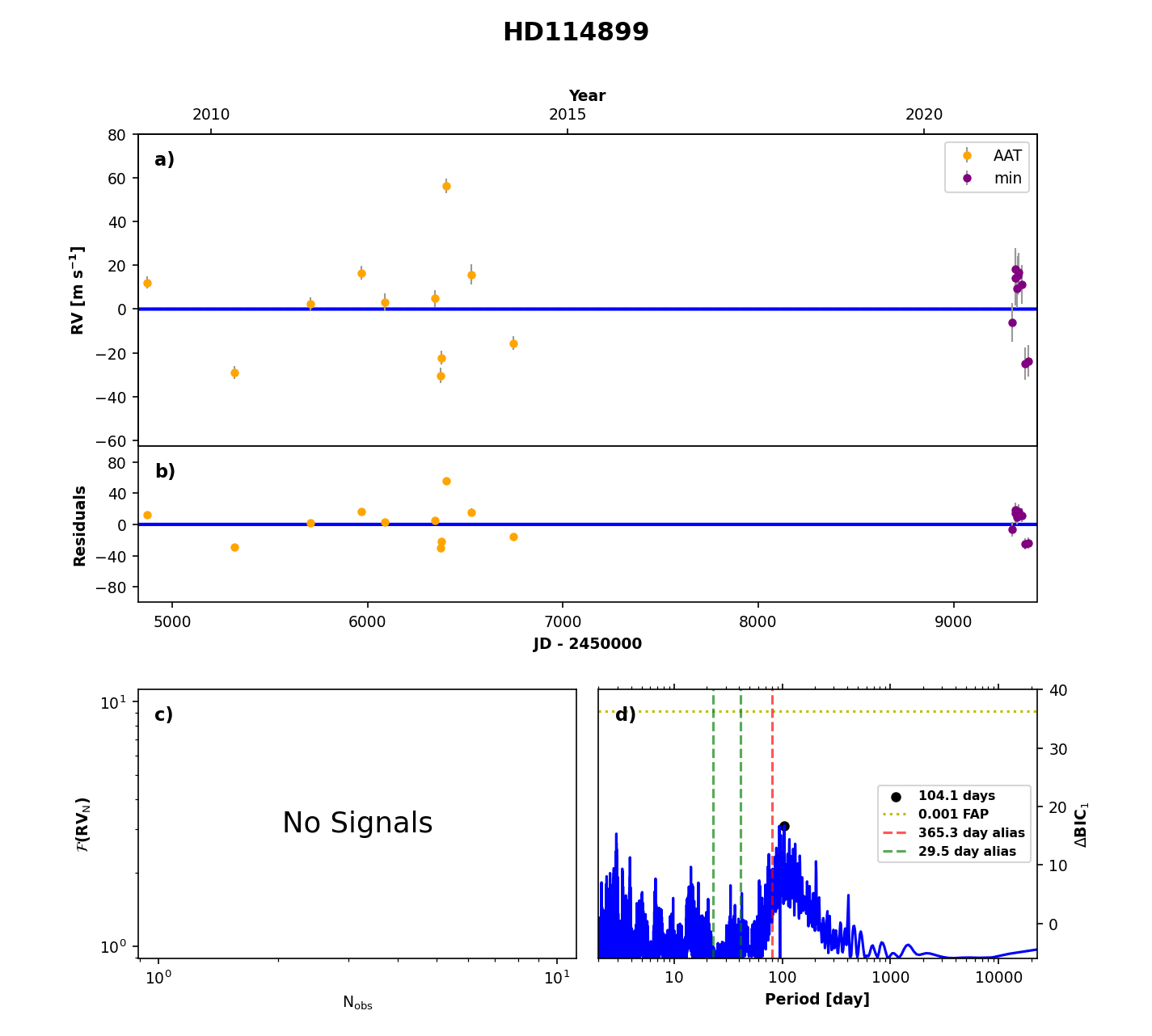}
     \includegraphics[width=\linewidth]{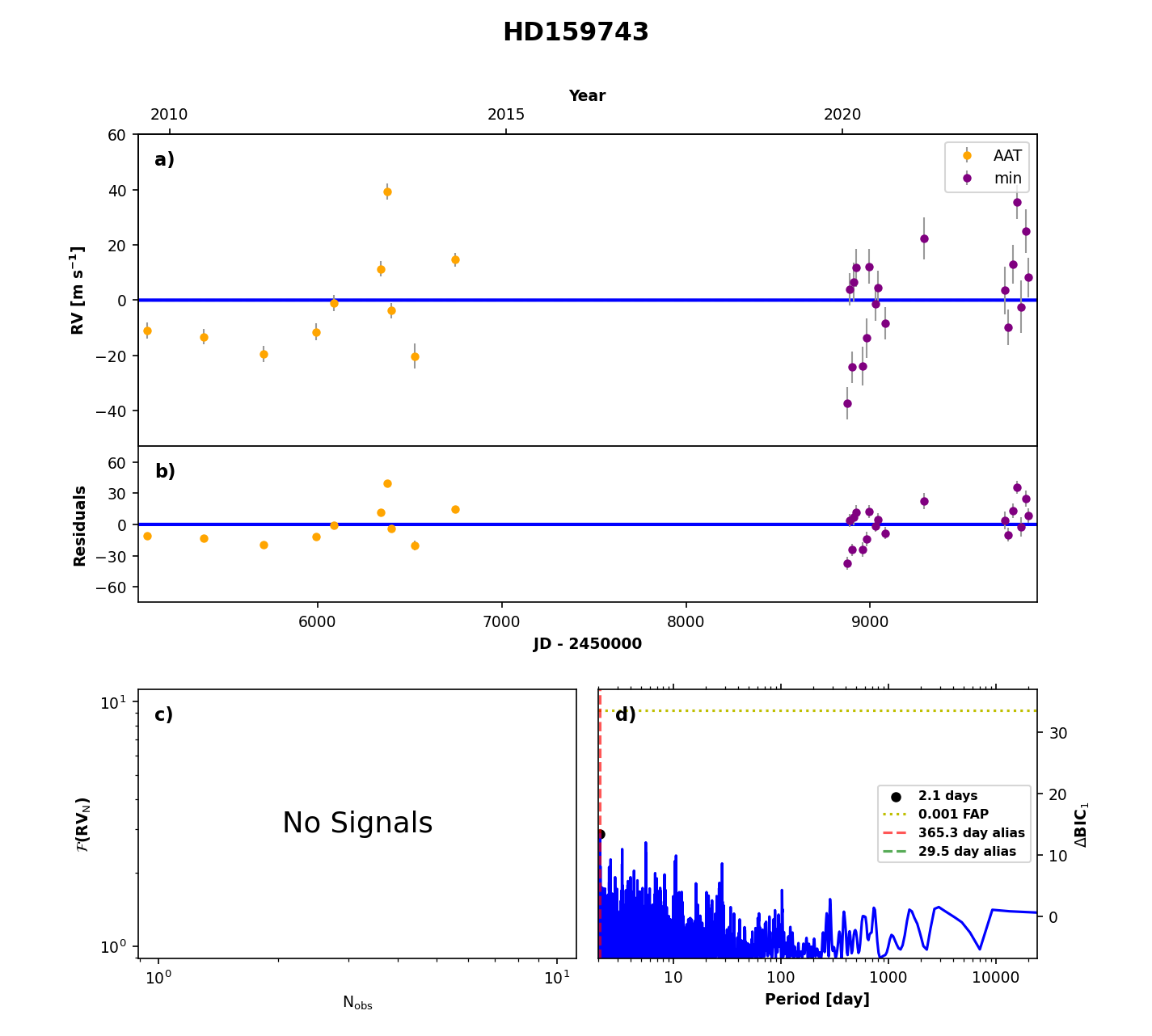}

     \caption{Results of a blind search in \texttt{RVSearch} for periodic signals in the combined data sets for HD\,114899 (top panel) and HD\,159743 (bottom panel).  No significant signals are found, thus dismissing the candidates suggested in \citet{ppps8}.  
     } 
    \label{notplanets} 
\end{figure}

\citet{ppps8} also proposed two candidate short-period Keplerian signals for HD\,114899 and HD\,159743.  We performed follow-up observations with \textsc{MINERVA}-Australis (9 epochs) for the former, and with SONG for the latter (19 epochs).  The new data for HD\,114899 did not support the proposed 42-day candidate planet: a Bayes Factor periodogram performed within the blind-search package \texttt{RVSearch} \citep{rosenthal21} shows no significant periodic signals whatsoever, ruling out the planet hypothesis.  The same analysis for HD\,159743's alleged 102-day planet gave a similarly conclusive repudiation of the planet model; both of these results are shown in Figure~\ref{notplanets}.  

\section{Discussion and Conclusions}\label{sec:Results}

We have followed up on six candidate companions that were tentatively proposed in \citet{ppps8} as part of the program's final data release.  Of these, we now confirm one giant planet, one high-mass brown dwarf, two stellar-mass companions, and we show the remaining two candidate signals to be spurious.  The companions orbiting HD\,115066 and HD\,121156 were presented in \citet{bluhm16} as likely stellar-mass objects moving on long, incomplete orbits.  Here we have analysed new RV data combined with Hipparcos-Gaia astrometric accelerations to confirm the nature of these objects as being low-mass stars.  Any discussion of radial-velocity detected companions must of course pay heed to the possibility of stellar activity.  The amplitudes of the signals we have confirmed here (cf. Table~\ref{tab:planets}) range from 37 to over 1300\,\ms.  By comparison, the scaling relations of \citet{Kjeldsen+Bedding1995} provide estimates of only $\sim$9-12\,\ms\ for the amplitudes of RV variations due to oscillations for these four evolved stars. 

HD\,126105b is a thoroughly ordinary member of the evolved-star planet population.  Indeed, its mass and eccentricity (1.7\mj\ and 0.10) are statistically indistinguishable from their median values (1.87\mj\ and 0.11) for the 417 planets known to orbit evolved stars\footnote{Planet data from NASA Exoplanet Archive, accessed 2025 Dec 28}.  HD\,205577B is a high-mass brown dwarf -- at its current best-fit mass of 77$^{+11}_{-9}$ \mj, it balances on the knife-edge of the hydrogen burning limit at 78.5\mj\ \citep{chabrier23}.  We note for future, more fortunate and prescient observers, that the next periastron passage of HD\,205577B is predicted for BJD 2461940, or 2028 June 17, though there remain uncertainties of several hundred days in period and $T_p$.  At that time, its eccentricity, and hence its true orbital period and mass, can be measured more precisely with a planned observational campaign.  This approach has proven vital in constraining high-eccentricity companions such as HD\,219077b \citep{marmier13}. HD\,45350b \citep{endl06}, HD\,20782b \citep{kane16}, HD\,76920b \citep{76920, bergmann21}, and HR\,5183b \citep{blunt19}. 

The eccentricities of giant planets, brown dwarfs and low-mass stars will shed light on whether systems with different mass ratios resemble each other, offering a unique way to trace their evolution histories. Recent work from \cite{Gan2025} looked into the eccentricity distributions of transiting long-period ($P>10$~days or $0.1\leq a\leq 1.5$~AU) giant planets, brown dwarfs and low-mass stars. The authors found that giant planets and brown dwarfs exhibit similar eccentricity distributions, preferring low-eccentricity orbits with a long tail toward high eccentricities. The low-mass stars, instead, tend to have moderately eccentric orbits with a peak at 0.3, a feature also found in massive stellar binaries \citep{Duquennoy1991,Wu2025}. However, wide-orbit companions seem to have different behaviors. Through the direct imaging method, several studies found that cold ($5\leq a\leq100$~AU) Jupiters and brown dwarfs have significantly different eccentricity distributions \citep{Bowler2020,Doo2023,Nagpal2023}. To explain such a discrepancy, \cite{Gan2025} proposed that giant planets and brown dwarfs probably form at exterior orbits through different channels, which result in different eccentricities, but then undergo an analogous evolution pathway, leading to similar eccentricity distributions of those inner systems. Therefore, enlarging the sample of systems with semi-major axis in-between ($1.5\leq a\leq 5$~AU), a parameter region that astrometry and RV are sensitive to, and investigating their eccentricity distributions will complement the picture, providing a more comprehensive insight.


\section*{Acknowledgements}

We respectfully acknowledge the traditional custodians of all lands throughout Australia, and recognise their continued cultural and spiritual connection to the land, waterways, cosmos, and community. We pay our deepest respects to all Elders, ancestors and descendants of the Giabal, Jarowair, and Kambuwal nations, upon whose lands the {\textsc{Minerva}}-Australis facility at Mt Kent is situated.

{\textsc{Minerva}}-Australis is supported by Australian Research Council LIEF Grant LE160100001, Discovery Grants DP180100972 and DP220100365, Mount Cuba Astronomical Foundation, and institutional partners University of Southern Queensland, UNSW Sydney, MIT, Nanjing University, George Mason University, University of Louisville, University of California Riverside, University of Florida, and The University of Texas at Austin.

Based on observations made with the Hertzsprung SONG telescope operated on the Spanish Observatorio del Teide on the island of Tenerife by Aarhus University and by the Instituto de Astrofísica de Canarias. 

This research has made use of the NASA Exoplanet Archive, which is operated by the California Institute of Technology, under contract with the National Aeronautics and Space Administration under the Exoplanet Exploration Program.

E.C. acknowledges support from the National Science Foundation under grant no. 1952545.

\textbf{Facilities}: {\textsc{Minerva}}-Australis; Stellar Observations Network Group 1m Hertzsprung Telescope. 

\textbf{Software}: RadVel \citep{radvelpaper}, RVSearch \citep{rosenthal21}, htof \citep{htof}, AstroImageJ \citep{2017AJ....153...77C}, isochrones \citep{morton15}, ispec \citep{ispec2014,ispec2019}, pyodine \citep{heeren23}

\section*{Data Availability}

The {\textsc{Minerva}}-Australis radial velocities underlying this article are available in the article.  The \textit{TESS} data used in this paper are also available via NASA's Mikulski Archive for Space telescopes:
\href{https://mast.stsci.edu/portal/Mashup/Clients/Mast/Portal.html}{https://mast.stsci.edu/portal/Mashup/Clients/Mast/Portal.html}



\bibliographystyle{mnras}
\bibliography{biblio, seismo, astrometry} 




\appendix

\section{Radial Velocity Data}

\begin{table}
	\centering
	\caption{Radial Velocities for HD\,114899}
	\label{tab:114899rv}
	\begin{tabular}{ccc}
		\hline
		Time & Velocity & Uncertainty \\
        BJD & \ms & \ms \\
		\hline
  \multicolumn3c{AAT/UCLES} \\
  2454870.195700  &       8.8  &    1.9  \\
  2455319.013170  &     -32.3  &    2.2  \\
  2455706.942220  &      -0.8  &    2.3  \\
  2455969.214640  &      13.3  &    2.3  \\
  2456088.970460  &       0.0  &    3.5  \\
  2456344.171560  &       1.6  &    3.2  \\
  2456375.098810  &     -33.6  &    2.9  \\
  2456378.055230  &     -25.6  &    2.3  \\
  2456400.157840  &      53.1  &    2.7  \\
  2456529.864310  &      12.5  &    4.1  \\
  2456746.205670  &     -18.8  &    2.4  \\
  \hline
\multicolumn3c{MINERVA-Australis pre} \\
  2459299.142001  &  -11578.5  &    8.7  \\
  2459315.290887  &  -11558.2  &   12.1  \\
  2459318.070336  &  -11553.9  &    9.4  \\
  2459324.003043  &  -11562.8  &    8.8  \\
  2459328.993926  &  -11557.0  &    8.5  \\
  2459332.043528  &  -11555.4  &    8.7  \\
  2459348.990869  &  -11560.9  &    8.6  \\
  2459365.898387  &  -11597.2  &    7.2  \\
  2459380.933720  &  -11596.0  &    7.0  \\
		\hline
	\end{tabular}
\end{table}

\begin{table}
	\centering
	\caption{Radial Velocities for HD\,115066}
	\label{tab:115066rv}
	\begin{tabular}{ccc}
		\hline
		Time & Velocity & Uncertainty \\
        BJD & \ms & \ms \\
		\hline
  \multicolumn3c{AAT/UCLES} \\
  2454870.212810  &    -152.0  &    2.8  \\
  2455317.978310  &    -254.3  &    1.6  \\
  2455706.957310  &    -129.2  &    2.2  \\
  2455757.891760  &    -110.0  &    5.9  \\
  2455787.886780  &    -123.0  &    3.5  \\
  2455908.236380  &     -60.5  &    2.2  \\
  2455969.225880  &      -6.6  &    2.0  \\
  2455995.191120  &       0.0  &    2.1  \\
  2456088.950390  &      29.0  &    2.5  \\
  2456344.195790  &     218.1  &    3.3  \\
  2456345.121980  &     220.1  &    3.0  \\
  2456376.112640  &     223.4  &    2.4  \\
  2456377.058730  &     237.5  &    3.1  \\
  2456528.851440  &     331.9  &    3.7  \\
  2456745.093680  &     515.9  &    2.3  \\
  \hline
  \multicolumn3c{MINERVA-Australis post} \\
  2460048.188247  &   -7550.1  &    4.6  \\
  2460066.150644  &   -7496.4  &    8.8  \\
  2460090.116868  &   -7475.8  &    5.2  \\
  2460112.004502  &   -7460.0  &    7.9  \\
  2460118.980981  &   -7442.6  &    6.6  \\
  2460125.914675  &   -7454.6  &    5.6  \\
  2460132.903789  &   -7452.1  &    6.6  \\
  2460137.875311  &   -7436.2  &    8.0  \\
  2460146.946522  &   -7419.5  &   11.7  \\
  2460152.854592  &   -7417.3  &    5.6  \\
  2460157.849174  &   -7417.7  &    7.3  \\
  2460176.853320  &   -7381.2  &    7.3  \\
		\hline
	\end{tabular}
\end{table}

\begin{table}
	\centering
	\caption{Radial Velocities for HD\,121156}
	\label{tab:121156rv}
	\begin{tabular}{ccc}
		\hline
		Time & Velocity & Uncertainty \\
        BJD & \ms & \ms \\
		\hline
  \multicolumn3c{AAT/UCLES} \\
   2454868.24554  &     874.1  &    1.3  \\
   2455227.21242  &     439.0  &    1.4  \\
   2455380.95382  &     241.6  &    1.6  \\
   2455580.25634  &      15.9  &    1.4  \\
   2455602.18474  &       3.0  &    1.9  \\
   2456060.06304  &    -251.4  &    2.5  \\
   2456090.96263  &    -252.4  &    1.8  \\
   2456345.14620  &    -255.6  &    1.5  \\
   2456376.21162  &    -233.1  &    1.7  \\
   2456527.87482  &    -147.8  &    3.7  \\
  \hline
  \multicolumn3c{MINERVA-Australis post} \\
  2460003.097372  &  -16594.3  &    5.2  \\
  2460008.300973  &  -16600.7  &    4.4  \\
  2460019.307448  &  -16590.0  &    4.4  \\
  2460045.195053  &  -16608.3  &    5.5  \\
  2460056.166556  &  -16589.1  &    4.1  \\
  2460062.931038  &  -16594.3  &    5.6  \\
  2460068.178250  &  -16579.0  &    4.6  \\
  2460073.932018  &  -16588.3  &    5.6  \\
  2460084.056714  &  -16588.1  &    4.3  \\
  2460097.887911  &  -16580.4  &    5.8  \\
  2460113.967355  &  -16577.7  &    5.7  \\
  2460118.991184  &  -16568.5  &    3.7  \\
  2460125.943859  &  -16574.1  &    4.2  \\
  2460136.894224  &  -16577.9  &    4.1  \\
  2460144.871345  &  -16565.5  &    4.9  \\
  2460149.859908  &  -16572.5  &    5.5  \\
  2460154.848652  &  -16567.6  &    4.9  \\
  2460159.857961  &  -16570.4  &    4.1  \\
  2460175.853850  &  -16571.5  &    4.5  \\
  2460186.850016  &  -16564.0  &    8.5  \\
  2460191.850487  &  -16582.6  &    7.4  \\
  2460352.121177  &  -16551.7  &    8.7  \\
  2460377.106745  &  -16538.0  &    7.3  \\
  2460386.054077  &  -16541.9  &    4.9  \\
  2460453.887000  &  -16550.7  &    4.9  \\
  2460478.019757  &  -16548.4  &    3.6  \\
  2460496.912360  &  -16550.3  &    4.8  \\
  2460550.850443  &  -16518.9  &   10.8  \\
  2460558.853096  &  -16514.6  &    5.1  \\
		\hline
	\end{tabular}
\end{table}

\begin{table}
	\centering
	\caption{Radial Velocities for HD\,126105}
	\label{tab:126105rv}
	\begin{tabular}{ccc}
		\hline
		Time & Velocity & Uncertainty \\
        BJD & \ms & \ms \\
		\hline
  \multicolumn3c{AAT/UCLES} \\
  2454868.253890  &     -16.4  &    1.9  \\
  2455380.999320  &     -10.1  &    2.1  \\
  2455581.241000  &      13.4  &    1.6  \\
  2455707.055580  &      21.8  &    1.9  \\
  2455969.271540  &     -14.3  &    1.4  \\
  2455994.133180  &       0.0  &    2.8  \\
  2456052.038380  &      28.4  &    4.2  \\
  2456060.037380  &      38.1  &    2.7  \\
  2456088.939590  &      34.8  &    1.9  \\
  2456344.210650  &     -39.1  &    2.1  \\
  2456375.293110  &     -46.8  &    1.7  \\
  2456377.140980  &     -49.7  &    2.1  \\
  2456400.070540  &     -39.4  &    2.0  \\
  2456494.886100  &     -13.4  &    2.2  \\
  2456746.229980  &      29.4  &    1.8  \\
  \hline
  \multicolumn3c{SONG} \\
  2458830.785861  &     521.7  &    5.1  \\
  2458858.716871  &     517.7  &    4.3  \\
  2458885.751962  &     509.9  &    3.8  \\
  2458914.741429  &     480.7  &    4.3  \\
  2458959.488011  &     459.7  &    5.2  \\
  2458977.442452  &     466.8  &    6.1  \\
  2458994.394525  &     478.9  &    4.3  \\
  2459042.451876  &     478.1  &    4.3  \\
  2459294.538720  &     563.1  &    6.8  \\
  2459717.383548  &     537.3  &    4.7  \\
  2459730.403288  &     528.0  &    6.7  \\
  2459747.471653  &     542.8  &    5.1  \\
  2459775.398244  &     527.5  &    5.2  \\
  \hline
  \multicolumn3c{MINERVA-Australis post} \\
2460380.049459   &  42254.8   &   7.5 \\
2460399.995394   &   42228.9  &  8.8 \\
2460414.991303   &  42235.5   &   5.7 \\
2460427.955808   &  42232.4   &  5.0 \\
2460436.932452   &   42231.0  &   7.9 \\
2460450.189079   &   42219.5  &   7.3 \\
2460453.139724   &   42215.8  &   5.7 \\
2460467.949451   &   42206.0  &   6.7 \\
2460475.133190   &   42223.8  &   5.8 \\
2460478.965789   &   42217.3  &   8.8 \\
2460485.017683   &   42203.1  &   7.1 \\
2460487.845079   &   42218.2  &  5.4 \\
2460495.921944   &   42203.3  &  5.4 \\
2460501.938646   &   42197.6  &    8.9 \\
2460505.900609   &   42185.9  &  10.1 \\
2460513.952257   &   42189.8  &  5.8 \\
2460518.858219   &   42193.4  &  6.3 \\
		\hline
	\end{tabular}
\end{table}

\begin{table}
	\centering
	\caption{Radial Velocities for HD\,159743}
	\label{tab:159743rv}
	\begin{tabular}{ccc}
		\hline
		Time & Velocity & Uncertainty \\
        BJD & \ms & \ms \\
		\hline
  \multicolumn3c{AAT/UCLES} \\
  2455075.026750  &      -7.2  &    2.0  \\
  2455382.072740  &      -9.5  &    1.8  \\
  2455707.227410  &     -15.8  &    2.1  \\
  2455994.274090  &      -7.7  &    2.3  \\
  2456089.086630  &       2.7  &    2.0  \\
  2456344.292410  &      15.1  &    1.9  \\
  2456377.242300  &      43.1  &    2.1  \\
  2456400.271850  &       0.0  &    2.0  \\
  2456528.972220  &     -16.5  &    4.0  \\
  2456746.286030  &      18.4  &    1.6  \\
  \hline
  \multicolumn3c{SONG} \\
2458876.792598  &  -75382.0  &    5.6  \\
2458886.768423  &  -75340.8  &    5.6  \\
2458899.761411  &  -75369.0  &    5.3  \\
2458910.736684  &  -75338.1  &    6.8  \\
2458921.706757  &  -75333.0  &    6.5  \\
2458959.731132  &  -75368.5  &    6.8  \\
2458982.673146  &  -75358.5  &    6.8  \\
2458994.641296  &  -75332.5  &    6.0  \\
2459027.626970  &  -75346.2  &    5.6  \\
2459041.385406  &  -75340.2  &    6.0  \\
2459079.374902  &  -75353.0  &    5.4  \\
2459292.693858  &  -75322.5  &    7.4  \\
2459730.527921  &  -75341.2  &    8.3  \\
2459747.503024  &  -75354.5  &    6.2  \\
2459775.431204  &  -75331.7  &    6.7  \\
2459795.437425  &  -75309.1  &    5.8  \\
2459817.376043  &  -75347.2  &    9.3  \\
2459843.415142  &  -75319.7  &    7.6  \\
2459856.335793  &  -75336.5  &    6.8  \\
		\hline
	\end{tabular}
\end{table}

\begin{table}
	\centering
	\caption{Radial Velocities for HD\,205577}
	\label{tab:205577rv}
	\begin{tabular}{ccc}
		\hline
		Time & Velocity & Uncertainty \\
        BJD & \ms & \ms \\
		\hline
  \multicolumn3c{AAT/UCLES} \\
  2455075.246820  &    -187.3  &    2.2  \\
  2455381.278790  &    -123.1  &    2.1  \\
  2455524.915130  &    -115.6  &    2.4  \\
  2456091.176550  &     -54.1  &    2.6  \\
  2456400.260490  &       0.0  &    2.6  \\
  2456494.161480  &      48.3  &    3.2  \\
  2456495.142060  &      34.6  &    2.2  \\
  2456529.099640  &      71.7  &    2.9  \\
  \hline
  \multicolumn3c{MINERVA-Australis pre} \\
  2459314.298135  &   -7303.4  &    5.2  \\
  2459324.278982  &   -7318.9  &    6.3  \\
  2459349.207404  &   -7283.2  &    5.7  \\
  2459364.327423  &   -7288.7  &    5.8  \\
  2459395.139557  &   -7294.7  &    4.5  \\
  2459421.291983  &   -7250.4  &    9.8  \\
  2459431.145560  &   -7278.0  &    4.5  \\
  2459447.042507  &   -7273.9  &    4.6  \\
  2459479.052971  &   -7283.4  &    6.0  \\
  2459504.886853  &   -7243.0  &    6.3  \\
  2459737.318410  &   -7217.0  &   22.4  \\
  2459810.030194  &   -7209.5  &    5.0  \\
  2459816.104450  &   -7193.6  &   16.1  \\
  \hline
  \multicolumn3c{MINERVA-Australis post} \\
2460068.204871  &   -7203.7  &    8.0  \\
2460072.197296  &   -7192.7  &    7.8  \\
2460073.195785  &   -7196.6  &   13.0  \\
2460074.177895  &   -7193.5  &    9.1  \\
2460075.221966  &   -7194.4  &    6.0  \\
2460076.172731  &   -7197.1  &    6.1  \\
2460077.288534  &   -7224.0  &    6.4  \\
2460078.167891  &   -7212.6  &    7.8  \\
2460081.285903  &   -7218.6  &    7.5  \\
2460083.160599  &   -7188.7  &    8.5  \\
2460085.162389  &   -7212.3  &    8.1  \\
2460087.160405  &   -7217.6  &   10.4  \\
2460088.156889  &   -7177.6  &   13.7  \\
2460090.135457  &   -7203.5  &   12.3  \\
2460091.133005  &   -7177.7  &   13.8  \\
2460093.128261  &   -7191.5  &   16.9  \\
2460094.137846  &   -7201.5  &   10.1  \\
2460095.123817  &   -7188.5  &    5.9  \\
2460096.177027  &   -7205.5  &   10.8  \\
2460097.118108  &   -7176.1  &    7.6  \\
2460105.127613  &   -7191.7  &    7.8  \\
2460109.123191  &   -7202.4  &    9.3  \\
2460110.121838  &   -7190.7  &    7.4  \\
2460111.119048  &   -7172.9  &   11.6  \\
2460112.112530  &   -7195.7  &    8.4  \\
2460114.104805  &   -7193.6  &    7.0  \\
2460115.102046  &   -7189.4  &    8.2  \\
2460116.165205  &   -7204.4  &    8.5  \\
2460119.090618  &   -7194.9  &   11.8  \\
2460120.118007  &   -7186.1  &   10.9  \\
2460121.115681  &   -7213.4  &    5.8  \\
2460126.085436  &   -7206.4  &   11.3  \\
2460131.058875  &   -7183.8  &    6.3  \\
2460132.101226  &   -7174.5  &    7.8  \\
2460133.053224  &   -7191.9  &    6.5  \\
2460134.050593  &   -7193.5  &    5.9  \\
		\hline
	\end{tabular}
\end{table}

\begin{table}
	\centering
	\contcaption{Radial Velocities for HD\,205577}
	\label{tab:205577rv_2}
	\begin{tabular}{ccc}
		\hline
		Time & Velocity & Uncertainty \\
        BJD & \ms & \ms \\
		\hline
  \multicolumn3c{MINERVA-Australis post} \\
2460135.335997  &   -7180.2  &    7.3  \\
2460136.045489  &   -7190.3  &    6.0  \\
2460137.043259  &   -7185.6  &    7.0  \\
2460140.315787  &   -7177.4  &    5.3  \\
2460142.029461  &   -7201.7  &    6.7  \\
2460145.053170  &   -7205.1  &    6.3  \\
2460155.086609  &   -7171.9  &    6.3  \\
2460170.997891  &   -7201.5  &    8.5  \\
2460465.209862  &   -7162.1  &    7.6  \\
2460485.088477  &   -7156.6  &    8.3  \\
2460506.040246  &   -7175.2  &   10.3  \\
		\hline
	\end{tabular}
\end{table}


\bsp	
\label{lastpage}
\end{document}